\documentclass[useAMS,usenatbib]{mn2e} 
\usepackage{amsmath,amssymb,wasysym,graphicx} 
 
\title[Near-Infrared Imaging Polarimetry of Young Stellar Objects in 
rho-Ophiuchi]{Near-Infrared Imaging Polarimetry of Young Stellar 
Objects in rho-Ophiuchi} \author[A. F. Beckford, P. W. Lucas, 
A. C. Chrysostomou and T. M. Gledhill]{A. F. Beckford, 
P. W. Lucas\thanks{E-mail: P.W.Lucas@herts.ac.uk}, A. C. Chrysostomou 
and T. M. Gledhill\\ Centre for Astrophysics Research, University of 
Hertfordshire, College Lane, Hatfield AL10 9AB,  United Kingdom\\} 
 
\begin{document} 
 
\date{Accepted 2007 . Received 2007 ; in original form Tuesday 17th 
April 2007} 
 
\pagerange{\pageref{firstpage}--\pageref{lastpage}} \pubyear{2008} 
 
\maketitle 
 
\label{firstpage} 
 
\begin{abstract} 
The results of a near-infrared (J H K L$_P$) imaging linear 
polarimetry survey of 20 young stellar objects in $\rho$ Ophiuchi are 
presented.  The majority of the sources are unresolved, with K band 
polarizations, P$_K$ $<$ 6\%.  Several objects are associated with 
extended reflection nebulae.  These objects have centrosymmetric 
vector patterns with polarization discs over their cores; maximum 
polarizations of P$_K$ $>$ 20\% are seen over their envelopes. 
Correlations are observed between the degree of core polarization and 
the evolutionary status inferred from the spectral energy 
distribution. K band core polarizations $>$6\% are only observed in 
Class I YSOs. 
 
A 3-dimensional Monte Carlo model with oblate grains aligned with a 
magnetic field is used to investigate the flux distributions and 
polarization structures of three of the $\rho$ Oph young stellar 
objects with extended nebulae. A $\rho$ $\propto$ r$^{-1.5}$ power law 
for the density is applied throughout the envelopes. The large scale 
centrosymmetric polarisation structures are due to 
scattering. However, the polarization structure in the bright core of 
the nebula  appears to require dichroic extinction by aligned 
non-spherical dust grains. The position angle  indicates a toroidal 
magnetic field in the inner part of the envelope. Since the measured 
polarizations attributed to dichroic extinction are usually $\le$10\%, 
the grains must  either be nearly spherical or very weakly 
aligned. The higher polarizations observed in the outer parts of the 
reflection nebulae require that the dust grains responsible for 
scattering  have maximum grain sizes $\leqslant$1.05~$\mu$m. 
\end{abstract} 
 
 
\section{Introduction} 
 
Multicolour imaging linear polarimetry at near-infrared (near-IR) 
wavelengths is a powerful tool for mapping the dusty discs and 
envelopes that surround young stellar objects (YSOs).  The observed 
flux distributions and the polarization patterns are both influenced 
by the distribution and properties of the dust grains responsible for 
the extinction and the scattering of the light. It can be used to 
determine which source in a given region dominates the illumination. 
It can also be used to distinguish between the different mechanisms 
that produce polarization, since dichroic extinction and scattering 
processes sometimes produce differing wavelength dependencies. 
 
Generally, the scattering pattern associated with spatially resolved 
YSOs is centrosymmetric with aligned vectors in the core of the 
nebula. This pattern of aligned vectors is commonly referred to as the 
polarization disc. Some YSOs with resolved nebulae differ from the 
traditionally expected centrosymmetric pattern by displaying a much 
broader region of aligned vectors. The most likely cause of this 
feature is dichroism due to magnetically aligned non-spherical dust 
grains in the circumstellar disc.  Other proposed methods for 
producing the aligned vector patter include multiple scattering 
(Whitney \& Hartman 1993) and ``illusory disc'', which is an effect 
caused by limited spatial resolution (Whitney, Kenyon \& Gomez 1997; 
Lucas \& Roche 1998). 
 
\begin{table*} 
 \begin{minipage}{150mm} 
    \caption{Rho Oph sample.} 
  \begin{tabular}{@{}lcccccccl@{}} 
  \hline Name\footnote{VSSG: Vrba, Strom, Strom \& Grasdalen 1975; 
   GSS: Grasdalen, Strom \& Strom 1973; DOAR: Dolidze \& Arakelyn 
   1959; WL: Wilking \& Lada 1983; EL: Elias 1978; YLW: Young, Lada \& 
   Wilking 1986; WLY: Wilking, Lada \& Young 1989 Table 2} & 
   \multicolumn{2}{c}{Coordinates (2000)}\footnote{Quoted from SMBAD} 
   & IR Class\footnote{Wilking, Lada \& Young 1989; Andre \& Montmerle 
   1994; Bontemps et al. 2001} & \multicolumn{4}{c}{Flux 
   (mags)\footnote{Wilking, Lada \& Young 1989; Greene et al. 1994; 
   Barsony et al. 1997}} & Alternative\footnote{EL: Elias 1978; YLW: 
   Young, Lada \& Wilking 1986; GY: Greene \& Young 1992; WL: Wilking 
   \& Lada 1983; WLY: Wilking, Lada \& Young 1989 Table 2; GWAYL: 
   Greene, Wilking, Andre, Young \& Lada 1994} \\ \cr & RA & DEC & 
   & J & H & K & L & Names\\ 
         
 \hline VSSG1  & 16 26 21.5 & -24 23 07 & II & 13.49 & 10.76 & 8.68 & 
 - & EL20, YLW31 \\  GSS30  & 16 26 21.5 & -24 23 07 &  I &13.89 & 
 10.83 & 8.32 & 6.1 & GY6, WL15 \\ DOAR25 &  16 26 23.7 & -24 43 13 & 
 II/D & 9.78 & 9.83 & 7.73 & 7.28 & GY17, YLW34 \\ WL16 & 16 27 02.5 & 
 -24 37 30 & II ? & 14.14 & 10.58 & 7.92 & 5.85 & GY182, YLW5A \\ EL29 
 & 16 27 09.6 & -24 37 21 & I & 17.21 & 12.01 & 7.54 & 3.88 & GY214, 
 YLW7A, WL15 \\ WL20 E & 16 27 15.9 & -24 38 46 & II & 13.54 & 10.78 & 
 9.21 & 8.62 & GY240, YLW11B \\ WL20 W & 16 27 15.69 & -24 38 43.4 & 
 II & - & - & - & - & - \\ WL3 & 16 27 19.3 & -24 28 45 & II & 
 $>$17.00 & 14.49 & 11.20 & 8.8 & GY249, WLY41, WLY41, YLW12D \\ 
 YLW13A & 16 27 19.5 & -24 41 40 & III & 9.42 & 8.69 & 8.41 & 8.22 & 
 GY250, WLY40, SR12 \\ YLW13B & 16 27 21.7 & -24 41 42 & II & 15.21 & 
 11.31 & 8.41 & 5.83 & GY252, WLY42 \\ WL6 & 16 27 21.8 & -24 29 55 & 
 I & $>$17.00 & 14.39 & 10.04 & 7.13 & GY254, YLW14A \\ WLY43 & 16 27 
 27.1 & -24 40 51 & I & $>$17.00 & 13.17 & 9.46 & 6.85 & GY265, 
 YLW15A\\ YLW16A & 16 27 28.3 & -24 39 33 & I & $>$17.00 & 13.09 & 
 9.65 & 6.68 & GY269, WLY44 \\ VSSG18 & 16 27 28.4 & -24 27 21 & II & 
 14.47 & 11.48 & 9.39 & & GY273, WLY45, EL32, YLW17A \\ YLW16B & 16 27 
 29.7 & -24 39 16 & I & $>$17.00 & 14.65 & 11.46 & 7.63 & GY274, 
 WLY46\\ WLY47 & 16 27 30.1 & -24 27 43 & II & 15.44 & 11.64 & 8.95 & 
 & GY279, EL33, YLW17B\\ WLY48 & 16 27 37.2 & -24 30 34 & I & 10.53 & 
 8.65 & 7.42 & 5.91 & GY304, YLW46 \\ WLY51 & 16 27 40.0 & -24 43 13 & 
 I & 17.12 & 12.42 & 8.93 & 6.15 & GY315, YLW45 \\ WLY54 & 16 27 51.7 
 & -24 31 46 & I &16.63 & 13.50 & 10.87 & 4.73 & GY378, YLW52 \\ WLY63 
 & 16 31 35.5 & -24 01 28 & I & 16.40 & 12.15 & 9.29 & 6.79 & L1709B, 
 GWAYL4 \\ WLY67 & 16 32 01.0 & -23 56 44 & I & $>$17.00 & 13.31 & 
 10.43 & 7.97 & L1689-GWAYL6 \\ \hline 
 
\end{tabular} 
\end{minipage} 
\end{table*} 
 
The $\rho$ Oph star-forming region, at a distance commonly quoted as 
160 parsecs, is one of the nearest sites of low-mass star 
formation. The cloud complex contains a number of distinct dark clouds 
(Lynds 1962) and filamentary clouds (or streamers), with a total mass 
estimated to be 10$^4$ M$_{\odot}$.  It has been studied extensively at 
wavelengths ranging from x-ray to radio (Sekimoto et al. 1997; Girart, 
Rodriguez \& Curiel 2000; Andre, Ward-Thompson \& Barsony 1993; 
Kamazaki et al. 2001).  It is known to contain a rich population of 
young stars associated with circumstellar envelopes and/or discs 
(Grasdalen, Strom \& Strom 1973; Vrba et al. 1975; Elias 1978; 
Wilking, Young \& Lada 1989; Greene \& Young 1992; Barsony et 
al. 1997; Bontemps et al. 2001; Wilking et al. 2001).  Current 
estimates based on infrared observations put the number of YSOs in the 
region at approximately 200 (Bontemps et al. 2001).  The proximity of 
$\rho$ Oph to our solar system and the wealth of identified YSOs make 
it a good site for a polarimetric study of the dusty discs and 
envelopes surrounding YSOs. 
 
In this paper we present the results of a near-infrared imaging 
polarimetry survey of young stellar objects in the $\rho$ Oph 
star-forming region.  We compare polarimetric data with assumptions 
about the evolution of YSOs to determine if there are any correlations 
between the degree of polarization observed and the evolutionary 
status determined from the IR SED, also looking for correlations with 
the object colour.  We also present grain scattering models for three 
of the Class I objects. 
 
In $\S$2 we discuss the observations, in $\S$3 we present the results 
of the observations, in $\S$4 we discuss the individual sources 
observed, in $\S$5 we present the computer models, and in $\S$6 we 
present the conclusions. 
 
\section{Observations} 
 
\begin{table*} 
    \begin{minipage}{150mm} 
    \caption{Rho Oph aperture linear polarimetry.} 
    \begin{tabular}{@{}lcccccccc@{}} 
  \hline Name & \multicolumn{4}{c}{P$_{core}$\footnote{ Evaluated in 2 
   arcsecond diameter apertures, except: GSS30, which is evaluated in 
   a 3 arcsecond aperture (to enable comparison with Chrysostomou et 
   al. 1996), and WLY47 arc which is evaluated in a 1 
   arcsecond aperture.  The quoted errors are based on the standard 
   deviation of the polarization within the aperture used to assess 
   the degree of polarization.}} & 
   \multicolumn{3}{c}{P$_{max}$\footnote{ Evaluated in 0.5 arcsecond 
   diameter apertures, the quoted errors are based on the standard 
   deviation within the aperture.}} & 
   $\theta_{disc}$\footnote{Position angle of the long axis of the 
   disc, evaluated from K waveband data where available }  \\ \cr & J 
   & H & K & L\footnote{Evaluated in 1 arcsecond diameter apertures} & 
   J & H & K  \\ 
         
 \hline VSSG1 &-&1.4$\pm$0.8&2.7$\pm$1.4&-&-&-&-&-\\ 
 VSSG1-NW&-&1.3$\pm$1.0&3.8$\pm$1.4&-&-&-&-&-\\ GSS30 
 &23.8$\pm$0.1&22.7$\pm$0.1&15.5$\pm$0.1&4.3$\pm$1.2&32.9$\pm$1.1&41.6$\pm$1.3&50.4$\pm$1.4&151.2\\ 
 DOAR25 &1.4$\pm$0.1&0.8$\pm$0.1&1.1$\pm$0.1&1.0$\pm$0.5& & &  &30.1\\ 
 WL16 &-&-&5.1$\pm$0.05\footnote{Shift \& Add Result}&-&-&-&-&33.9\\ 
 EL29 &-&-&8.3$\pm$0.06$^e$&-&-&-&35.3$\pm$0.6$^e$&28.3\\ WL20 E 
 &11.5$\pm$1.2&-&-&-&-&-&-&38.0\\ WL20 W 
 &12.5$\pm$1.2&-&-&-&-&-&-&33.7\\ WL3 &-&-&2.1$\pm$0.2&-&-&-&-&20.3\\ 
 YLW13A &-&0.98$\pm$0.05 &0.8$\pm$0.05&-&-&-&-&165.3\\ YLW13A S 
 &-&9.7$\pm$0.7&-&-&-&-&-&172.9\\ YLW13B 
 &-&10.2$\pm$0.2&6.3$\pm$0.2&4.4$\pm$0.1$^e$&-&-&-&11.6\\ WL6 
 &-&-&3.3$\pm$0.5&-&-&-&-&135.5\\ WLY43 
 &-&3.2$\pm$1.4&3.7$\pm$1.8&1.8$\pm$0.9&-&-&-&-\\ 
 WLY43-NW&-&4.9$\pm$2.1&2.1$\pm$1.05&-&-&-&-&-\\ YLW16A 
 &-&11.4$\pm$0.3&10.4$\pm$0.2&3.1$\pm$1.1& 
 &21.9$\pm$1.3&20.8$\pm$1.2&32.6\\ VSSG18 
 &16.2$\pm$7.0&3.1$\pm$0.9&2.1$\pm$1.1&8.7$\pm$1.2&37.2$\pm$2.4&22.2$\pm$2.2&19.5$\pm$2.9&145.0\\ 
 YLW16B &-&9.7$\pm$0.7&6.2$\pm$0.4&5.7$\pm$0.3&-&-&-&53.8\\ WLY47 
 &-&5.2$\pm$0.02&4.6$\pm$0.01&-&-&-&-&179.1\\ WLY47 ''arc'' 
 &-&11.3$\pm$0.03&7.7$\pm$0.02&-&-&28.3$\pm$2.1&36.1$\pm$2.2&179.1\\ 
 WLY48 &-&-&2.3$\pm$0.05$^e$&0.9$\pm$0.3&-&-&-&133.5\\ WLY51 
 &-&9.1$\pm$0.2&5.1$\pm$0.1&1.6$\pm$0.7&-&-&-&23.4\\ WLY54 
 &9.2$\pm$0.05&5.3$\pm$0.01&2.4$\pm$0.01&4.3$\pm$0.01&-&30.6$\pm$2.1&43.2$\pm$2.3&2.1\\ 
 WLY63 &-&8.0$\pm$0.07&3.9$\pm$1.8&-&-&-&-&150.1\\ WLY67 
 &-&9.5$\pm$2.2&2.2$\pm$1.0&-&-&39.4$\pm$3.2&24.1$\pm$5.3&0.3\\ \hline 
\end{tabular} 
\end{minipage} 
\end{table*} 
 
The sample contains 18 of the sources identified as Class I, based on 
the shape of their IR SEDs and spectral indices, from the Wilking, 
Lada \& Young (1989) IR survey of the $\rho$ Oph region.  Subsequent 
investigation combining sub-millimetre data with the IR data has led 
to 9 of the sources being re-classified as Class II (Andre \& 
Montmerle 1994).  It is these latter classifications that are 
typically quoted by other authors and are therefore adopted by this 
paper.  The Class II sources were retained in the sample to allow 
comparisons to be made. 
 
In addition, data has been obtained for a further 2 objects, the Class 
III object YLW13A and the Class II object WLY47.  Table 1 is a 
complete list of all 20 of the sources, their coordinates, IR 
classifications, and J, H, K and L magnitudes. 
 
The observations were made at the United Kingdom Infrared Telescope 
(UKIRT) in Mauna Kea, Hawaii during the nights of 1998 June 17-18, 
1999 April 28-May 01, 2000 July 01-03 and 2002 May 15-17.  The 
instrument used was IRCAM with the polarimeter module IRPOL2, designed 
and built at the University of Hertfordshire.  The J (1.2 $\mu$m), H 
(1.6 $\mu$m) and K (2.2 $\mu$m) broadband filters were used during the 
1998, 1999 and 2002 sessions, and the L$_P$ (3.8 $\mu$m) band filter 
was used during the 2000 session.  The instrument optics provided a 
plate scale of 0.286 arcseconds per pixel (0.143 arcseconds with the 
magnifier).  This gave a typical field of view of 36$\times$8 
arcseconds when using the magnifier.  Prior to the 2000 observing 
session IRCAM was upgraded, providing a plate scale of 0.0814 
arcseconds per pixel and a field of view of approximately 20$\times$4 
arcseconds.  During the 2002 observing run UFTI replaced IRCAM3/TUFTI, 
providing a plate scale of 0.091 arcseconds per pixel and a field of 
view of 90$\times$15 arcseconds.  Typically the seeing was between 0.5 
and 1.0 arcsec. 
 
The Wollaston prism mounted in IRPOL2 splits the radiation into the 
orthogonally polarized beams, usually referred to as the ordinary and 
extraordinary (o- and e-) beams for historical reasons. A focal plane 
mask is used to reduce the field of view into two strips, each about 
260$\times$50 pixels squared, to prevent the o- and e- beams from 
overlapping. The half-waveplate in the system is successively rotated 
through 0, 45, 22.5 and 67.5 degrees. The advantage of this technique 
is that the intensity of both the o- and e- beams are measured 
simultaneously, which improves the reliability of the data by greatly 
reducing the effects of variations in atmospheric transparency and 
seeing, allowing any variations in atmospheric transparency to be 
accounted for. 
 
To remove bad pixels on the array a 3-point jitter pattern mosaic, 
with 5 arcsecond east-west offsets, was used. 
 
\subsection{Shift \& Add Data} 
 
Four of the sample sources, EL 29, WL 16, IRS 54 and IRS 48, were 
imaged in the K-band using the shift and add image-sharpening 
technique. This technique involves the taking of very short exposures, 
which are shifted so that their peak pixels coincide and are then 
co-added. This is all performed in real-time by the ALICE (Array 
Limited Control Electronics) electronics system of IRCAM3. The 
magnifier was adopted during the shift and add runs providing a pixel 
scale of 0.143 arcseconds. 
 
\begin{table*} 
 \begin{minipage}{140mm}  
    \caption{Rho Oph source properties.} 
    \begin{tabular}{@{}lccl@{}} 
  \hline Name & $\alpha_{IR}$\footnote{Wilking, Lada, Young 1989; 
   Bontemps et al. 2001} & Polarization Vector& Total Intensity 
   Distribution \cr & & Pattern & \\ 
         
 \hline VSSG1  & -0.49 & -& Point-like\\ BKLT\footnote{The 2nd source 
 in the VSSG1 field has been identified as possibly being BKLT 
 J162618-242618} & -& -& Point-like\\ J162618-242818 & & & \\  GSS30 
 & 1.20 & Centrosymmetric& Highly extended bipolar nebula\\ DOAR25 & 
 -1.58 & Random& Point-like\\ WL16 & 0.79 & Aligned& Point-like \\ 
 EL29 & 0.98 & Centrosymmetric& Point-like, polarized intensity 
 images\\ & & & - possibly faint bipolar nebulosity \\ WL20 E & -0.07 
 & Aligned& Point-like, member of a triple system \\ WL20 W & -0.07 & 
 Aligned& Point-like, member of a triple system \\ WL3 & 0.23 & 
 Aligned& Point-like \\ YLW13A & -2.48 & Random& Point-like, in the H 
 band there is a\\ & & & possible second source to the south\\ YLW13A 
 S & -& Random& Point-like \\ YLW13B & 0.08 & Aligned& Point-like\\ 
 WL6 & 0.59 & Aligned& Point-like\\ WLY43 & 0.98 & -& Point-like 
 (slightly oval)\\ GY263\footnote{The 2nd source in the WLY43 field 
 has been identified as possiblly being GY263} & - & - & Point-like\\ 
 YLW16A & 1.57 & Centrosymmetric& Extended, bipolar nebula \\ VSSG18 & 
 -0.24 & Centrosymmetric& Highly extended cometary nebula \\ YLW16B & 
 0.94 & Aligned& Point-like \\ WLY47 & 0.17 & Aligned& Point-like, arc 
 of nebulosity to the NW  \\ WLY47 arc & -& Centrosymmetric & 
 Elongated, curved structure\\ WLY48 & 0.18 & Random& Point-like \\ 
 WLY51 & -0.04 & Aligned& Point-like, possible close\\ & & & binary 
 companion \\ WLY54 & 1.76 & Centrosymmetric& Extended, cometary 
 nebula \\ WLY63 & 0.4 & Aligned& Point-like \\ WLY67 & 0.74 & 
 Centrosymmetric& Extended, cometary nebula \\ \hline 
\end{tabular} 
\end{minipage} 
\end{table*} 
 
The main advantage of the shift and add observing technique is in the 
resolving of compact bipolar nebulae and small polarization discs. 
However, there are several major limitations of the method.  The data 
are read noise limited due to the short integration times involved, 
which means that it is much less sensitive than the conventional 
background limited polarimetry.  It is not possible to select or 
discard frames in a given stack.  The method can only be used to 
observe YSOs that have a prominent compact flux peak.

The source WLY54 was also imaged during the 1998 UKIRT run. 
Comparisons of the results of both imaging methods reveals that the 
structure of the total flux distributions and the polarization vector 
patterns is comparable. 
 
\subsection{ Data Reduction} 
 
The initial reduction of the data was performed using the Starlink 
software package {\sc CCDPACK}. Each image was dark subtracted and then 
divided by a suitable normalized flatfield. Our flatfields for J, H 
and K are median filtered images constructed from images of the sky 
taken at each half-waveplate position. For the L band data the 
flatfields are median filtered images constructed from source images 
at different offsets. The sky subtraction and the extraction of the o- 
and e- beam images was performed using the Starlink software package 
POLPACK. 
 
The images were combined with the same software to generate the Stokes 
parameters, I, Q and U, 
 
The polarization is 
 
\begin{equation} 
P=\frac{\sqrt{Q^2+U^2-\sigma^2}}{I}\\ 
\end{equation} 
 
\noindent{where $\sigma^2$ is the variance on Q or U. The position 
angle of the polarization is} 
 
\begin{equation} 
\theta =0.5\arctan \left( \frac{U}{Q}\right).\\ 
\end{equation} 
 
\section{Results} 
 
\begin{figure*}
\vspace{-4.6cm}
\hspace{-1cm} \includegraphics[scale=0.8,angle=0]{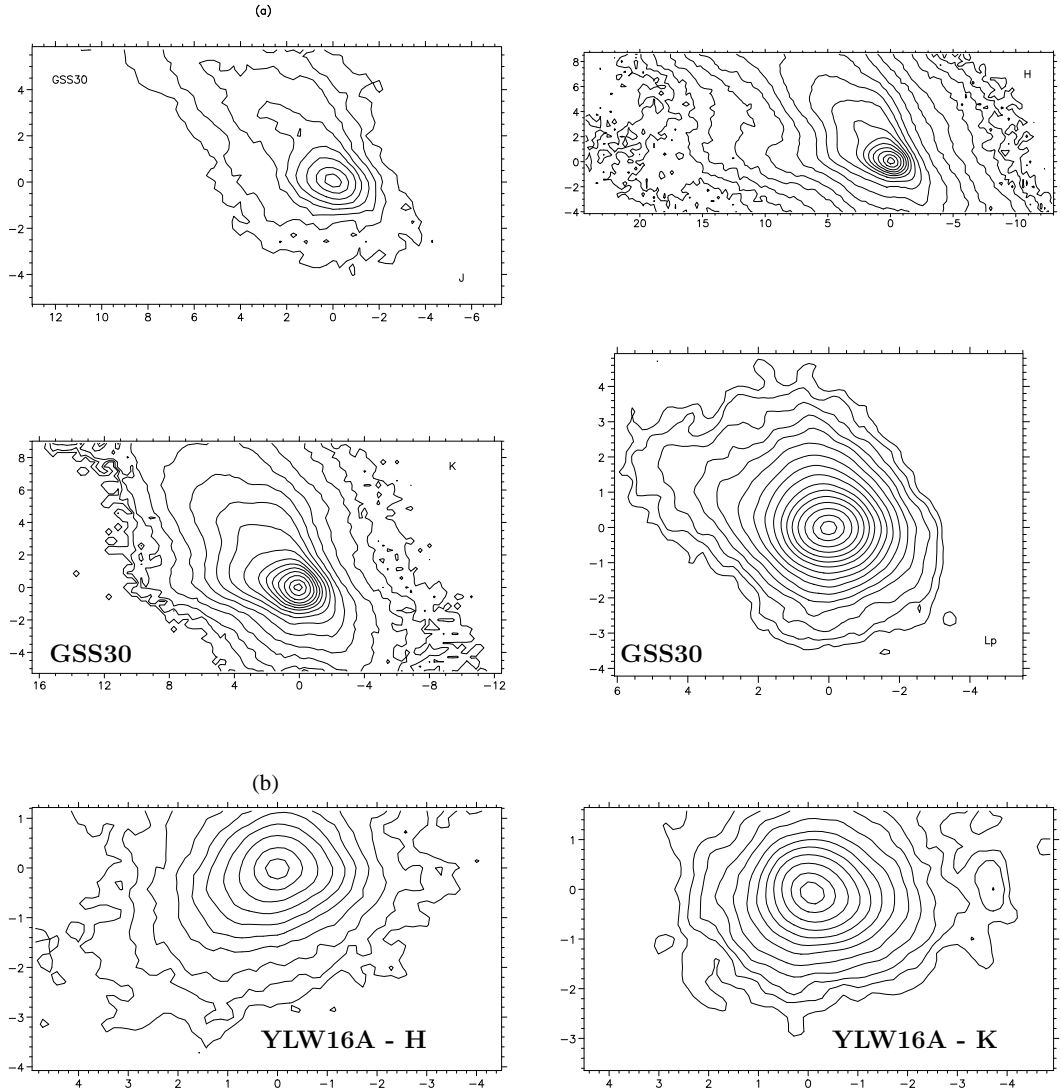} 
\vspace{-3cm} 
 \caption{Contour plots of the flux distributions for (a) GSS30, (b) 
YLW16A. Both GSS30 and YLW16A are bipolar nebulae, although the bipolar nature 
of YLW16A is only apparent in previously published higher resolution images (see 
text) and in the polarised flux and degree of polarisation maps (see Figure 4). 
Contours are normalised and spaced at 
0.9,0.71,0.5,0.35,0.25,0.18,0.13,0.088,0.006,0.004,0.003,0.002.} 
\end{figure*} 
 
\setcounter{figure}{0} 
\begin{figure*} 
 \vspace{-4.6cm} 
\hspace{-1cm} \includegraphics[scale=0.8,angle=0]{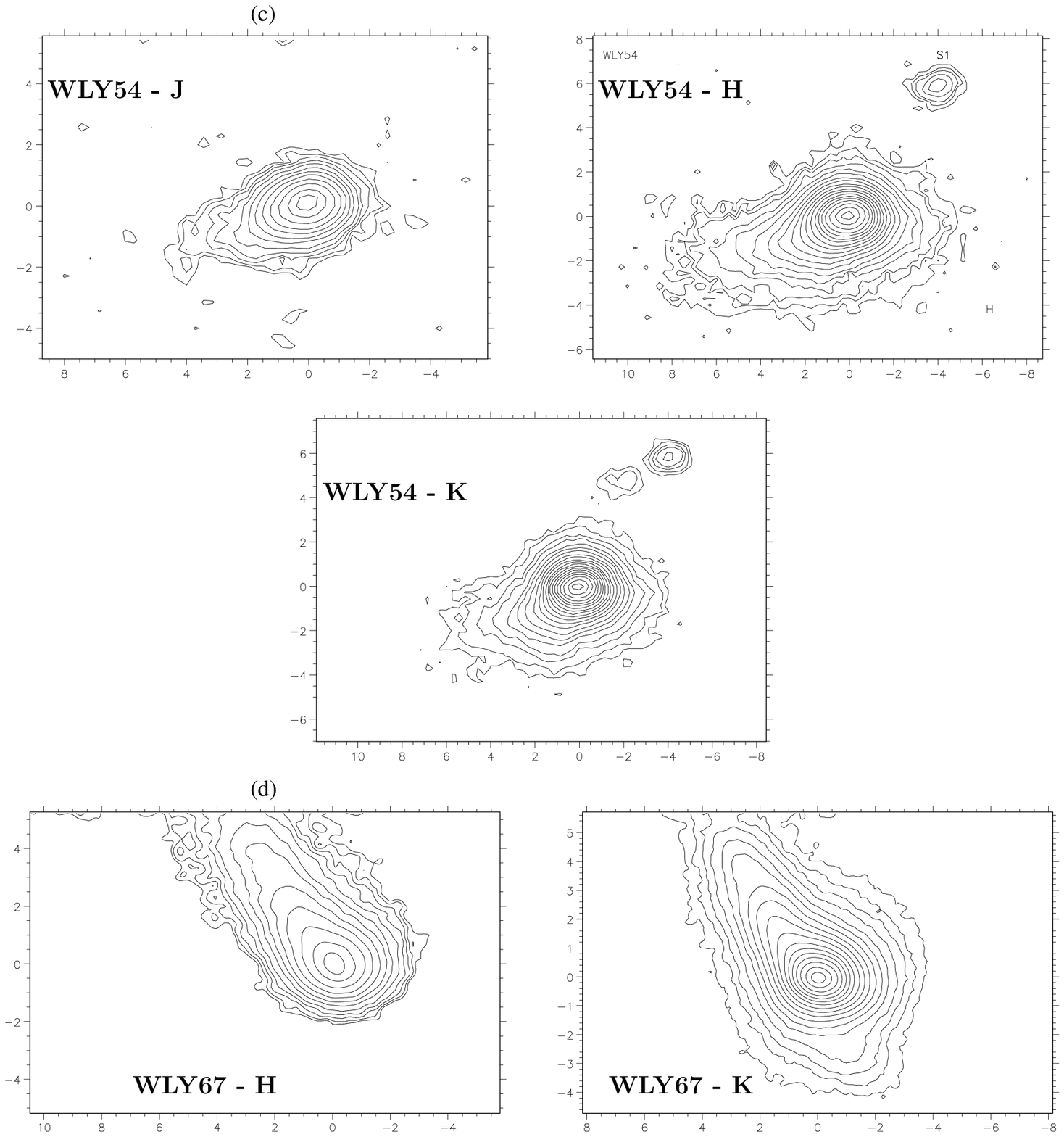} 
\vspace{-3cm} 
\caption {(c) WLY54, (d) WLY67. Both these YSOs display cometary nebulae. 
Contours are as in Figure 1(a-b)} 
\end{figure*} 
 
\setcounter{figure}{0} 
\begin{figure*} 
 \vspace{-4.6cm} 
\hspace{-1cm} \includegraphics[scale=0.8,angle=0]{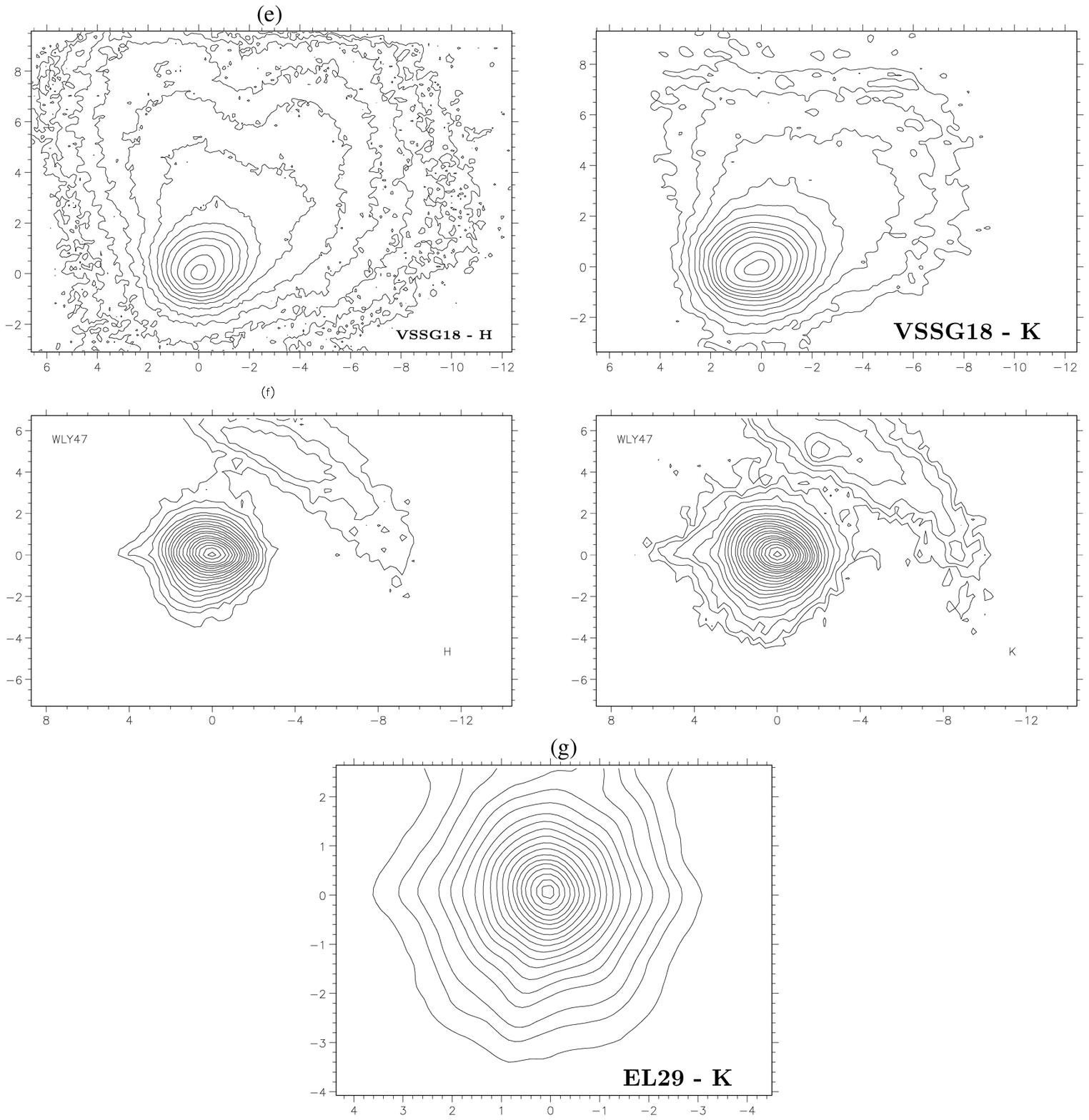} 
\vspace{-3cm} 
\caption {(e) VSSG18, (f) WLY47, and (g) EL29. VSSG18 is a cometary nebula, while 
WLY47 is a point source which appears to illuminate a detached arc of nebulosity. 
EL29 is unresolved in this image of the total flux but is resolved in polarised 
flux (see Figure 8). Contours are as in Figure 1(a-b)} 
\end{figure*} 
 
The majority of the sources are point-like, displaying no obvious 
extended structure.  It is possible that these objects may be 
associated either with compact or faint nebulosity that the detecting 
instrument is not sensitive to.  Five of the sources are clearly 
associated with extended nebulosity.  Two of these appear to have 
bipolar flux distributions (GSS30 and YLW16A), i.e. they have two 
lobes of extended nebulosity (evidence for YLW16A having two lobes of 
extended nebulosity can be seen in flux distribution maps that cover 
more of the region than can be seen in the polarization data presented 
here).  The remaining objects appear to have cometary morphologies 
(WLY54, WLY67 and VSSG18), i.e. there appears to be only one lobe of 
extended nebulosity and any counterlobe that may be present is not 
visible.  In addition to these five objects, the Class II source WLY47 
appears to be associated with a nebulous arc a few arcseconds away 
from the point source. 
 
The flux distributions of the five extended objects and WLY47 are 
shown in figures 1(a)--(g). 
 
\subsection{Polarization} 
 
The results of aperture polarimetry on the 20 sources are presented in 
Table 2.  The core polarizations are evaluated in 2 arcsecond 
apertures, centred on the flux peak.  The maximum polarizations are 
evaluated using 0.5 arcsec apertures, the exact positioning of the 
centre of these apertures is source dependent but for each source the 
same position is used at each wavelength.  Maximum polarizations are 
only shown for sources that are either clearly associated with 
extended nebulosity, or, in the case of EL29, show a significant 
spatial variation in fractional polarization within the image, which 
is attributed to small scale nebulosity that is hidden in the wings of 
the image profile.  Where there is more than one infrared object in 
the field of view, the degree of polarization of each is shown 
separately.  In total, aperture polarization results are presented for 
25 objects. 
 
The polarization vector patterns observed are divided into three 
categories dependent on their appearance:  centrosymmetric, aligned, 
and random.  The pattern is said to be centrosymmetric when the 
polarization vectors are arranged in a circle (or ellipse) about the 
main illuminating source.  When the polarization vectors appear to be 
arranged in ``parallel lines'' the pattern is aligned.  In a random 
pattern there is no apparent structure to the arrangement of the 
polarization vectors, indicating that no polarization was detected. 
In Table 3 the spectral index, $\alpha$, for each source is presented 
with a description of both its polarization vector pattern and its 
structure in the total flux image (i.e. the Stokes I parameter). 
 
\section{Individual Sources} 
 
\subsection{Compact Objects} 
 
The Class I objects YLW16B, WL6, WLY43, WLY48, WLY51 and WLY63, the 
Class II objects VSSG1, DOAR25, YLW13B, WL3, WL16 and WL20, and the 
Class III object YLW13A all have point-like morphologies in both total 
flux and polarised flux.  Several of these objects were found to be 
associated with a companion.  There is a second object located 
approximately 1120 AU from the core of WLY43, which has been 
tentatively identified as GY263, the structure of the companion is 
point-like.  There is a small knot approximately 2 arcseconds (320 AU) 
to the north of the core of WLY51 at a position angle of approximately 
20$^{\circ}$, which is surrounded by a horn of material.  The 
appearance of the knot is stronger in the H band, and is only 
suggested by the K band data.  There is a second object to the west of 
VSSG1.  The second object has been identified as possibly being the 
infrared source BKLT J162618-242818.  WL20 has been identified as a 
triple source by previous authors (Ressler \& Barsony 2001).  In our 
polarimetry data two of the 
members of the system are visible.  These are the Class II objects 
WL20 east (WL20 E) and WL20 west (WL20 W).  There is a bridge of 
material linking both sources. 
 
The degree of polarization over the cores of the Class I objects range 
from P$_H$ $\sim$ 9\% and P$_K$ $\sim$ 6\%, for YLW16B, to the very 
low value of P$_K$ $\sim$ 2\%, for WLY48.  The levels of core 
polarization for the Class II objects are at the lower end of those 
for the Class I objects. 
 
DOAR25 and WL3 only show levels of $\it{P}$$_K$ $\sim$ 1-2\%, whereas 
the levels observed for YLW13B in the H and K wavebands are 
$\it{P}$$_H$ $\sim$ 10\% and $\it{P}$$_K$ $\sim$ 6\%.  The spectral 
index for DOAR25, $\alpha_{2-14}$ $\sim$ -1.58, is the lowest 
displayed by any of the Class II objects; in some classification 
schemes it would be considered a Class III object.  The spectral index 
for YLW13B, $\alpha_{2-14}$ $\sim$ 0.08, identifies it as a possible 
transition object, the relatively high levels of polarization would 
seem to support this.  Almost all the compact objects display aligned 
polarization vectors, the exception being WLY48 which has a random 
vector pattern. YLW13A is the only Class III source that was imaged. 
The polarization vectors are best described as being randomly 
orientated, with an average position angle of $\theta_K$ $\sim$ 
165$^{\circ}$.  A 1-$\sigma$ upper limit on the fractional 
polarization  of $<$1\% is measured in both the H and K wavebands.  A 
second source is visible to the south of YLW13A in the H band.  The 
degree of polarization over YLW13A south is $\it{P}$$_H$ $\sim$ 9\%. 
 
\subsection{WLY47} 
 
WLY47 displays an interesting structure (see Figure 1f).  WLY47 
appears to be point-like; the interesting feature is the 'arc' of 
nebulosity that can be seen to the northwest.  The arc extends from 
the north to the west of the source and is approximately twelve 
arcseconds (2000 AU) in length.  There is evidence of a concentration 
of flux near the centre of the arc in both the H and K wavebands, 
possibly indicating that the arc contains a 2nd object. The core of 
WLY47 is approximately 5.5 arcseconds (880 AU) from the peak of the 
arc.  There is nebulosity surrounding both WLY47 and the arc that is 
apparent at low signal to noise.  HST NICMOS images also reveal the 
presence of the arc close to WLY47; the arc lies between WLY47 and 
VSSG18 (Allen et al. 2002). 

\begin{figure} 
 \vspace{0cm}\hspace{-2.8cm}\includegraphics[scale=0.9,angle=0]{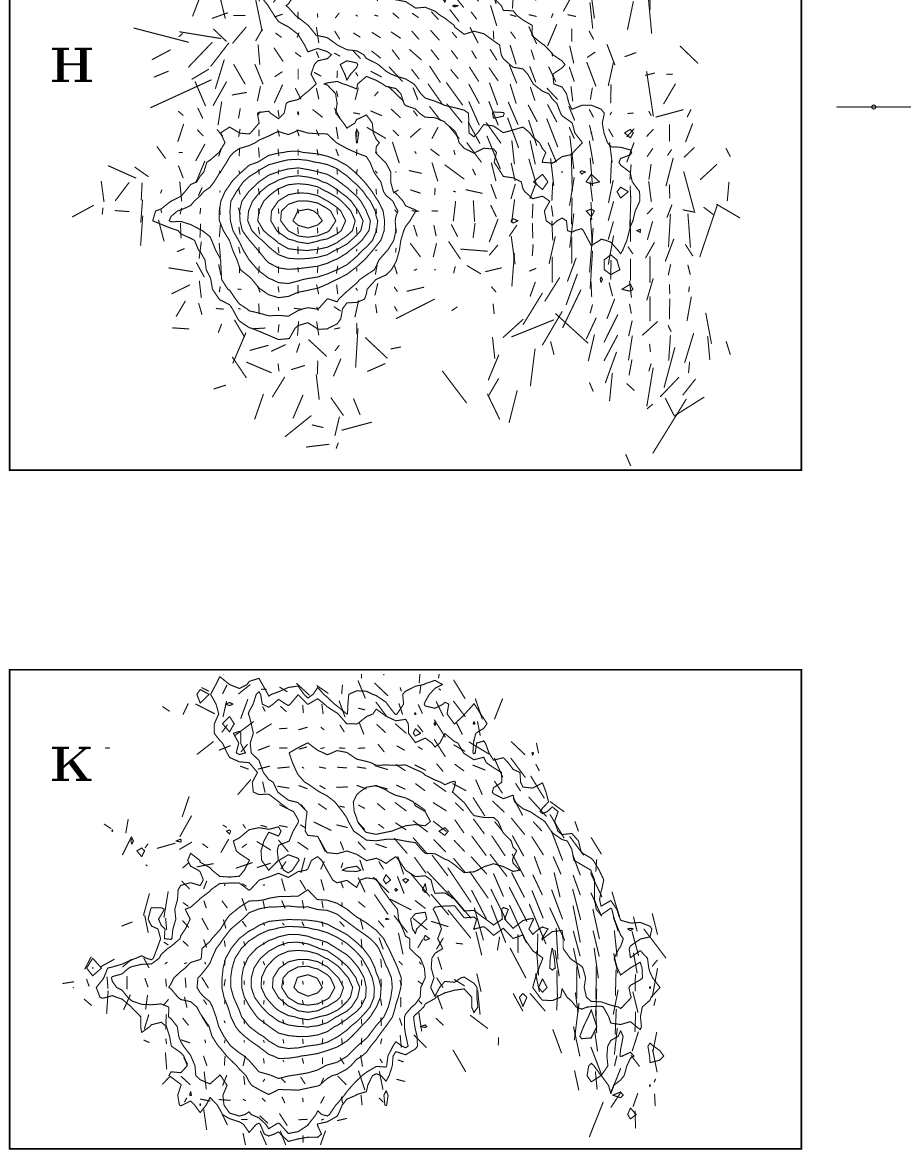} 
\vspace{-13.4cm} 
 \caption{WLY47 polarization. Contour plots of the flux distributions 
with polarization vectors overlaid for WLY47 in the H (upper) and K 
(lower) bands. The arc of nebulosity displays a centrosymmetric polarisation 
pattern, demonstrating that it is illuminated by the adjacent bright YSO.  
The key is the same length as a 100\% vector.} 
\end{figure} 
 
Figure 2 shows the H and K band polarization vector maps for WLY47. 
At both the H and K wavebands the polarization vectors across WLY47 
are aligned, whilst the vectors over the `arc' follow its structure, 
in the sense that the polarization vectors over the `arc' appear to be 
centrosymmetric with respect to WLY47. This indicates that the arc is 
a reflection nebula illuminated by WLY47.  The position angle of the 
polarization vectors over WLY47 and the peak of the arc are $\theta_K$ 
$\sim$ 179$^{\circ}$  and $\theta_K$ $\sim$ 50$^{\circ}$, 
respectively.  The levels of polarization observed are higher 
($\it{P}$$_H$ $\sim$ 5\% and $\it{P}$$_K$ $\sim$ 4\%) than those 
displayed by the majority of the Class II sources observed.  The 
levels of polarization over the peak of the arc of nebulosity are 
higher than those seen over the core of WLY47 in both wavebands, 
($\it{P}$$_H$ $\sim$ 11\% and $\it{P}$$_K$ $\sim$ 8\%.  The 
polarization increases toward the southern end of the arc. This may 
indicate that the arc does not lie in the plane of the sky and that 
scattering angle is closer to 90$^{\circ}$ at the southern end.  The 
maximum levels of polarization given in Table 2 are evaluated over the 
`arc' and are shown to be $\it{P}$$_H$ $\sim$ 28\% and $\it{P}$$_K$ 
$\sim$ 36\%. 
 
\subsection{GSS30} 
 
GSS30 has a bipolar morphology, with a northeast-southwest 
orientation, see Figure 1a and  Figure 3. This structure is 
interpreted as a reflection nebula, with the bright lobes 
corresponding to reflection from the walls of a bipolar cavity in the 
circumstellar nebula, which is assumed have been cleared by a bipolar 
outflow.  The northern lobe is brighter and more extensive than its 
southern counterpart.  This has been previously used by Chrysostomou 
et al. (1996) to estimate the inclination of the system.  Their 
investigation indicated that the system is inclined at an angle of 
approximately 25$^{\circ}$ - 30$^{\circ}$ to the plane of the sky, 
with the northern lobe tilted towards us.  Near-IR data covering more 
of the GSS30 region has revealed an extensive bipolar nebula that 
contains three distinct sources (Chrysostomou et al. 1996).  The main 
illuminating source is given the designation IRS1; the other two 
sources are IRS2 and IRS3, both lie in the northern lobe of nebulosity. 
 
The polarization vector map is shown in figure 3(a).  The polarization 
vector pattern is centrosymmetric in the outer regions, becoming more 
elliptical towards the main illuminating source (IRS1).  There is 
evidence of a narrow polarization disc over the core of IRS1 with a 
position angle $\theta_K$ $\sim$ 151$^{\circ}$, at the J, H, K and 
L$_P$ wavebands.  The polarization disc runs perpendicular to the 
orientation of the bipolar extension.  There is no apparent offset 
between the position of the disc and the central flux peak.  To the 
south of IRS1, along the polarization disc, the vectors experience a 
90$^{\circ}$ flip in their orientation, returning to the 
centrosymmetric pattern.  This return to the centrosymmetric vector 
orientation is not observed at the northern end of the disc in the J, 
H and K waveband maps, a feature previously noted by Chrysostomou et 
al.(1996).  In the L$_P$ band map a return to centrosymmetric pattern 
is seen at both ends of the polarization disc. These reversals are 
predicted in Monte Carlo multiple scattering models that do not 
include dichroic extinction effects (e.g. Figure 12 of Lucas \& Roche 
1998). The return to a centrosymmetric pattern at the ends of the 
polarization disc is caused by the transition from the optically thick 
part of the disc to the optically thin region where single scattering 
dominates the pattern, as opposed to multiple scattering. It is likely 
that the absence of this reversal at the northern end of the disc in 
the shorter wavelength data is due the greater extinction at shorter 
wavelengths. Greater extinction would increase the size of the 
optically thick, multiple scattering region and it would also 
strengthen any dichroic extinction effects which may also be 
contributing  to the observed polarisation disc, if the magnetic field 
has a toroidal structure along this line of sight. 
 
\begin{figure*} 
   \vspace{-5.8cm} 
 \hspace{-3cm}
\includegraphics[scale=1.0,angle=0]{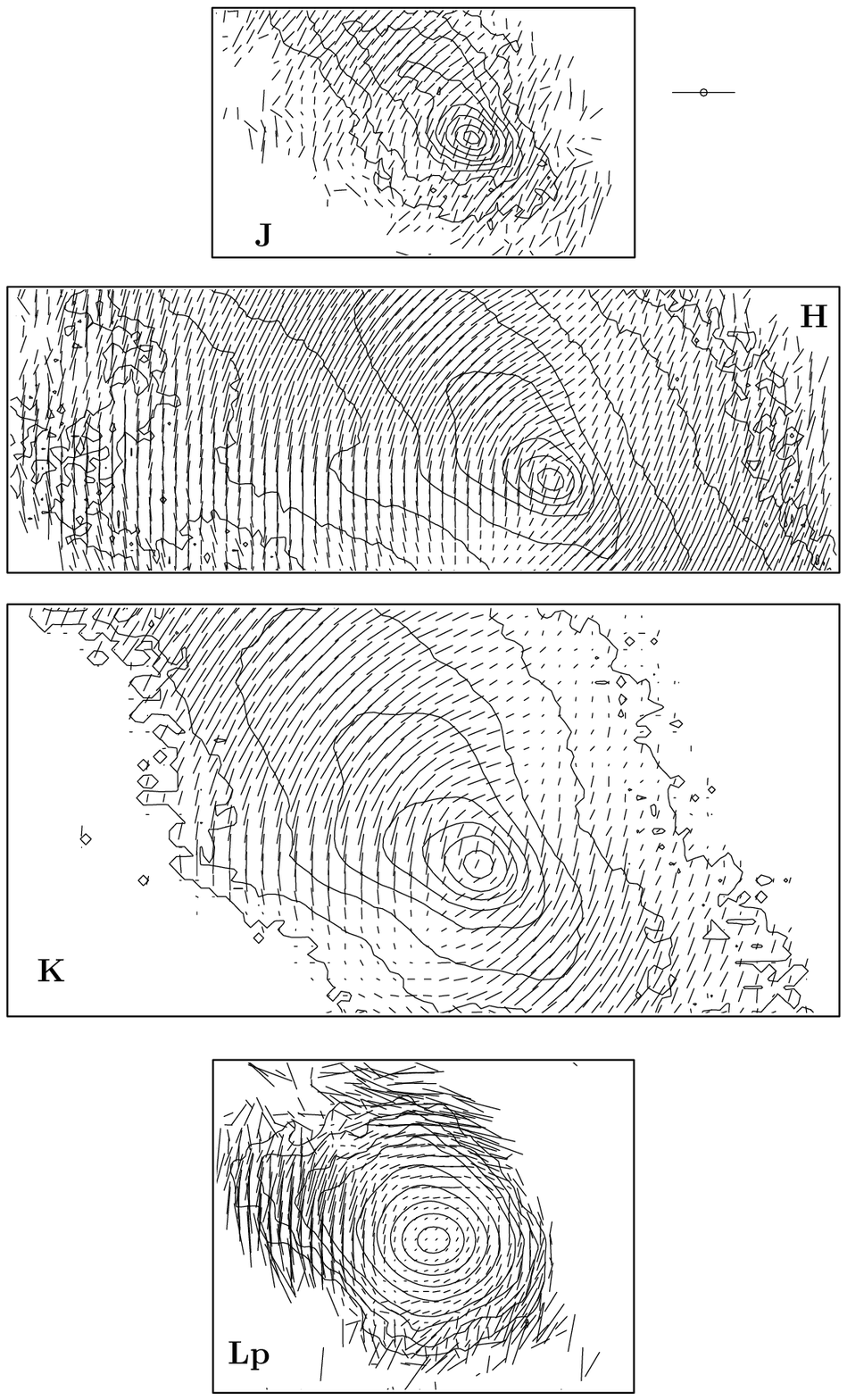}\vspace{-5.6cm} 
  \caption{GSS30 polarization. (a) Contour plots of the flux 
  distributions with polarization vectors overlaid in the J (top), H 
  (upper middle), K (lower middle), and L$_P$ (bottom) bands (the key 
  is the same length as a 110\% vector).} 
\end{figure*}  
 
\setcounter{figure}{2} 
\begin{figure} 
 \vspace{-1cm}\hspace{-1.3cm}\includegraphics[scale=0.6,angle=0]{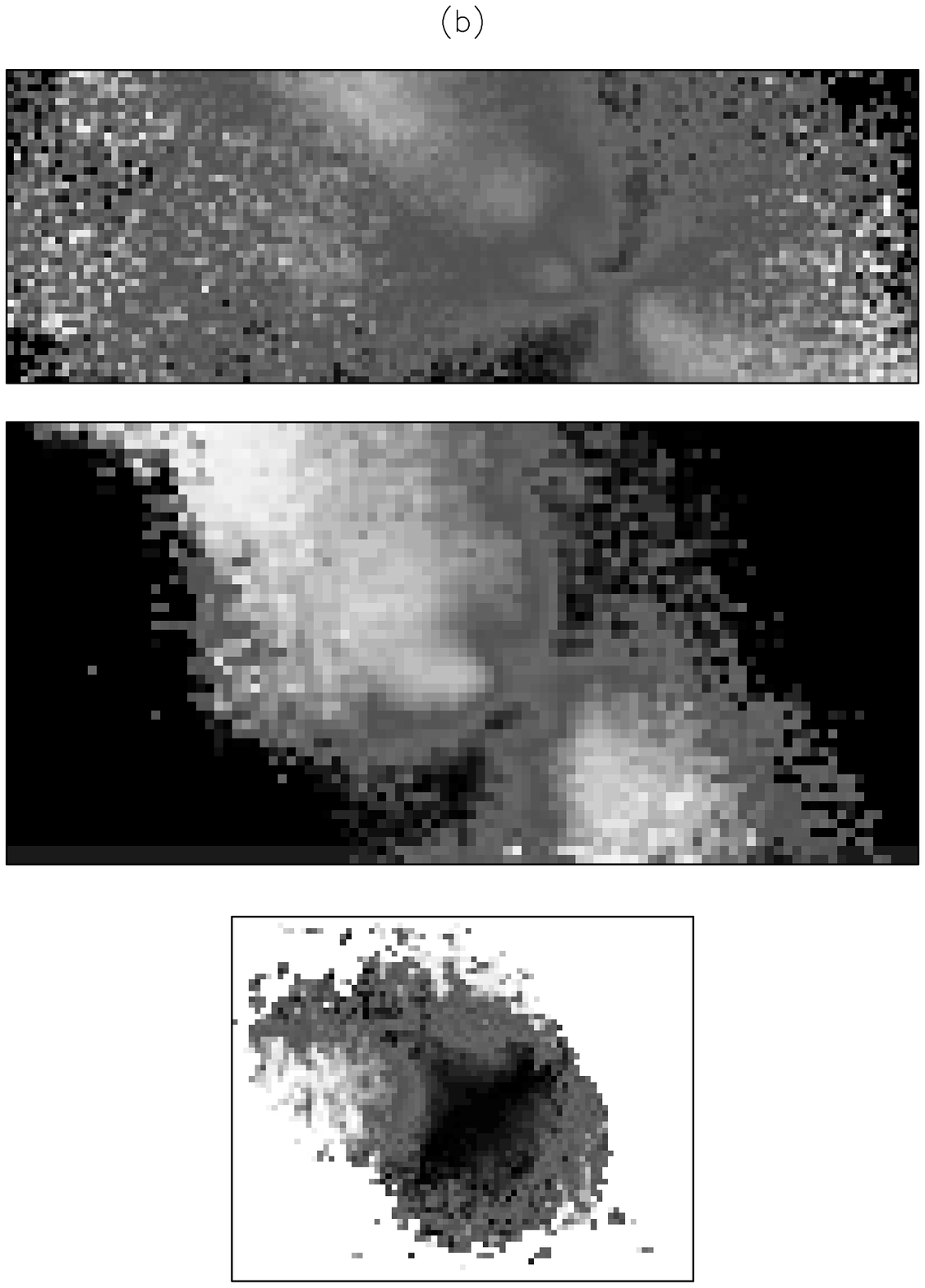}\vspace{-1.8cm} 
\caption{(b). Degree of polarization maps for GSS30. The panels are for the H (top),  
K (middle), and L$_P$ (bottom) bands.} 
\end{figure}  
 
\setcounter{figure}{2} 
\begin{figure} 
 \vspace{-1cm}\hspace{-1.3cm}\includegraphics[scale=0.6,angle=0]{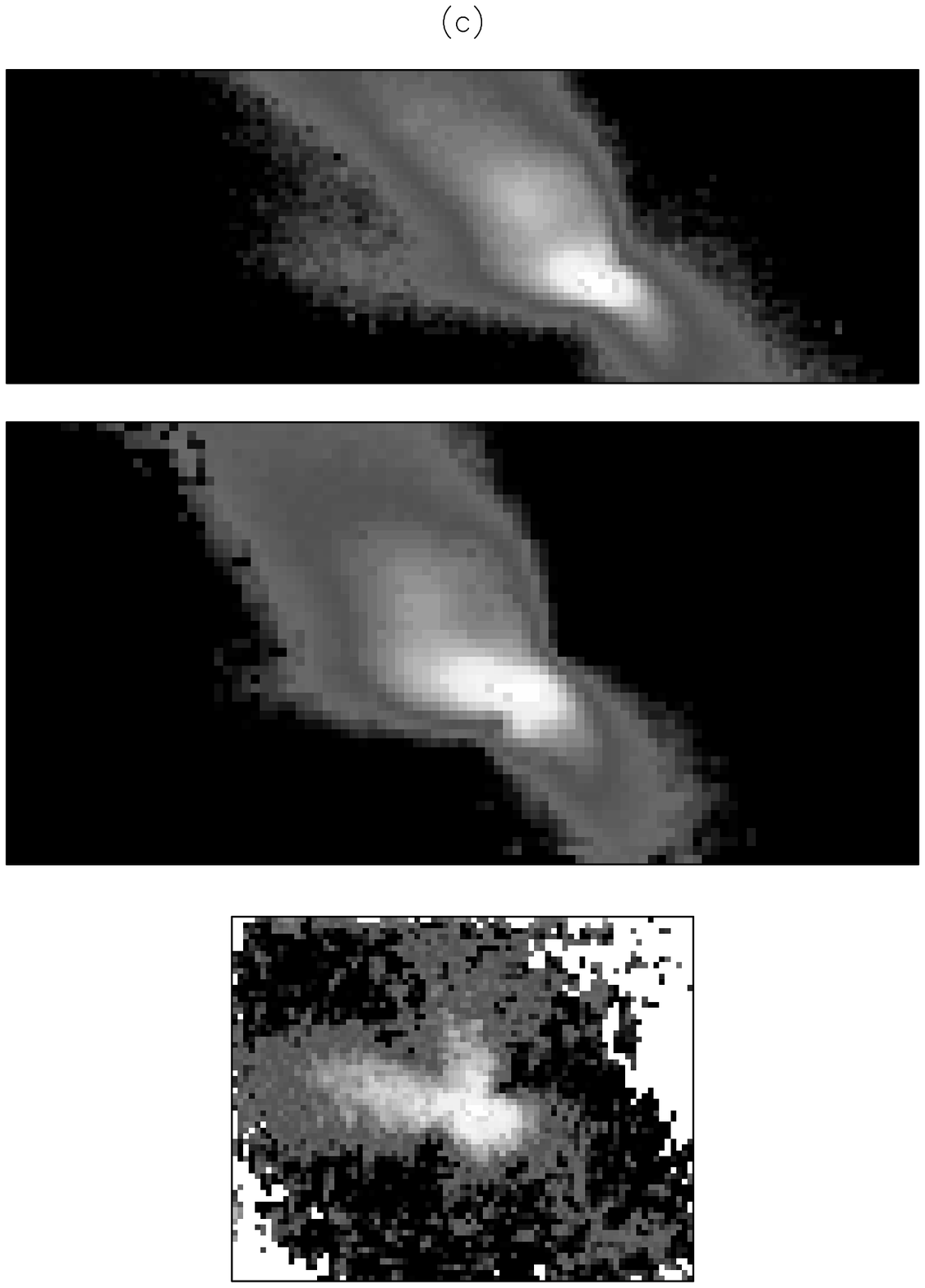}\vspace{-1.8cm} 
\caption{(c). Maps of the polarized flux for GSS30 in the H (top), K (middle), and 
 L$_P$ (bottom) bands.} 
\end{figure}  
 
In figures 3(b) and (c) images of the degrees of polarization and 
polarized flux at the H, K and L$_P$ bands are shown, respectively. 
(Polarized flux in each pixel is the product of flux and degree of  
polarisation). The H, K and L$_P$ band degree of polarization maps have  
a lower polarization region representing the polarization disc.  The low 
polarization region is surrounded by a higher polarization region, 
which marks the extent of the nebulosity shown in figure 3(a).  In 
the L$_P$ band map there is also a lower polarization region that 
extends in the direction of the polarization disc axis, for the length 
of the bright north lobe.  This region is narrower than the low 
polarization region that marks the polarization disc.  A similar 
feature is not seen in either the H or K band maps.  The L band 
observations penetrate low density matter and any knots of dust in the 
bipolar cavity to show the influence of the dense material more 
clearly. It is likely that the low polarization region seen at L$_P$ 
band along the polarization disc axis is simply due to the smaller 
scattering angle for material  in the walls of the bipolar cavity that 
is projected along the axis, compared to off-axis  locations. At 
shorter wavelengths this effect (previously seen in IRAS 04302+2247 by 
Lucas \& Roche 1997) may be obscured by low density matter within the 
cavity. Future studies of GSS30 should look at the object at radio and 
millimetre wavelengths to determine the appearance of the disc,
see Zhang, Wootten \& Ho (1997).
 
In the L$_P$ band polarized flux map, there appear to be three lobes. 
The smallest is the southern lobe seen in the flux distribution.  The 
larger northern lobe that is apparent in the total flux distribution 
is seen as two lobes in the polarized flux map. We interpret this as  
a simple consequence of the region of low polarization located between  
the two lobes. In the H and K 
band maps (figure 3(c)) the structure revealed is similar to that seen 
in the flux distribution but the flux distribution appears pinched 
in the disk plane, revealing the bipolar nature of GSS30 more clearly. 
This is due to the low polarization in the disk plane, caused by multiple 
scattering in this optically thick region. There are two lobes that extend  
northeast and southwest. Close to IRS1, where the polarized light is 
brightest, the lobes appear to extend more along an east-west axis.   
This inner east-west extension is not seen in the total flux distribution, 
but it does roughly align with the horn of nebulosity that extends to the 
east of the northern lobe. The eastern extension in polarized flux 
is due to a region of high polarization shown in figure 3(b). This may 
be caused by a region of low extinction inside the cavity to the east of  
IRS1, which would minimse the amount of depolarization due to  
multiple scattering. 
 
\begin{figure}
  \vspace{-2cm}\hspace{-1.6cm}\includegraphics[scale=0.6,angle=0]{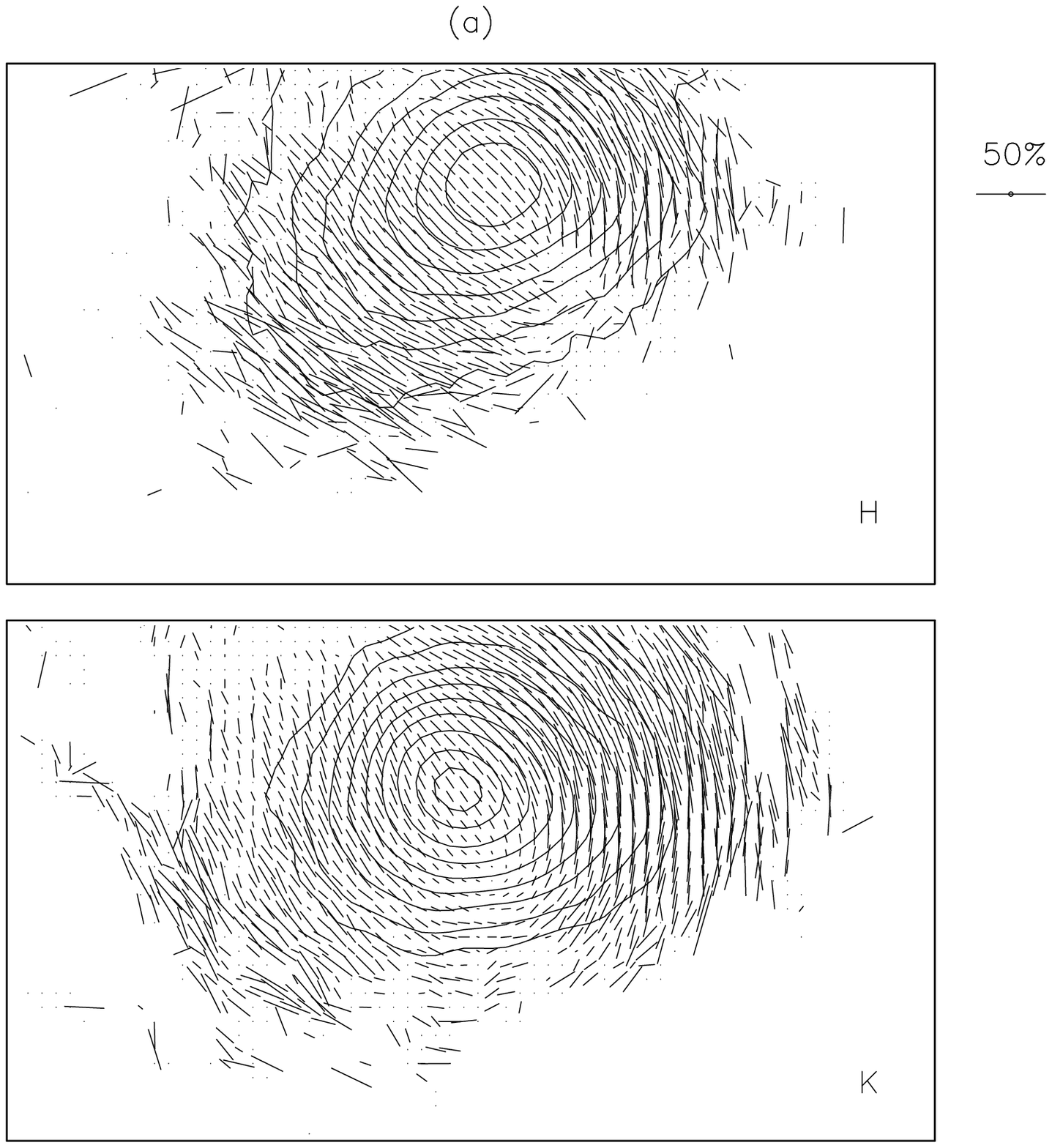}\vspace{-1.8cm} 
 \caption{YLW 16A polarization. (a) H and K band contour plots of the 
 flux distributions with polarization vectors overlaid, the key is the 
 same length as a 50\% vector, (b) Degree of Polarization maps, and 
 (c) Polarized flux distribution maps for YLW16A.  The data are (b) 
 scaled between 2\% (black) and 45\% (white), and (c) on a logarithmic 
 scale between 1.6 (black) to 2.5 (white) at H and 1.6 (black) to 3.1 
 (white) at K.} 
\end{figure}  
 
\subsection{YLW16A} 
 
The data presented covers the central 7$\times$5 arcseconds 
(1125$\times$800 AU) of the YLW16 region. The structure revealed by 
the polarization data in Figure 4 is a bipolar nebula,  with an 
approximately east-west extension; in the H band the inner contours 
have a more northwest-southeast orientation. The bipolar structure is 
less obvious from the Stokes I contours  (see Figure 1b) but it is clear 
in the higher resolution HST NICMOS image shown in figure 15 (see also
Allen et al.2002). Previous authors have shown that the nebula surrounding 
YLW16A has a radius of approximately 3400 AU (Aspin, Casali \& Walther 
1989; Lucas \& Roche 1998).  Lucas \& Roche (1998) commented that the 
outer contours at H-band have an interesting structure reminiscent of 
an arrowhead.  At K-band they found that this structure was less 
prominent. The HST NICMOS images of the source reveal two non-point sources 
at the position of YLW16A.  
The resolved nature of these sources make it difficult to determine 
whether there are actually two sources, or one source being seen in scattered 
light. Their appearance at 1.1 and 1.6 $\mu$m suggests that there is 
only one source. 
 
The polarization vector pattern is centrosymmetric in the outer 
regions.  In the core region there is a broad polarization disc which 
is far more extended along the axis of the bipolar nebula than is 
typical and is evident in both the H and K band maps (see figure 
4(a)). The strength of this feature strongly implies that it is 
produced by dichroic extinction by grains whose short axis is 
preferentially aligned with the disc plane., since multiple scattering 
models with spherical grains do not reproduce it, see section 
5.3. Furthermore, the position angle of polarisation of the adjacent 
unresolved source YLW16B is very similar to that of the polarization 
disc in YLW16A, which also suggests that dichroic extinction is 
operating.

In order to test this hypothesis we removed the dichroic component by subtracting 
the average Q/I and U/I values for YLW16B from the Q/I and U/I frames for YLW16A. The 
resulting effect on the polarization vector pattern of YLW16A is that the H band  
polarization disc is not as broad as shown in figure 3(a), and it has a position 
angle of $\theta_H$ $\sim$ 0$^{\circ}$.  In the K band the 
polarization disc is no longer clearly visible. This is what would be expected
if dichroic extinction is primarily responsible for the polarization disk, so
the evidence supports this interpretation.

There is no apparent offset between the polarization disc and the 
central flux peak.  The polarization disc is also observed to be 
approximately perpendicular to the direction of the inner contours, 
with a position angle of $\theta_K$ $\sim$ 32$^{\circ}$ and $\theta_H$ 
$\sim$ 40$^{\circ}$.  This is the largest variation in the position 
angle of the polarization disc between wavebands shown by any of the 
sources.  In the K band polarization vector map, following the length 
of the polarization disc to the south the vectors return to the 
centrosymmetric pattern.  The same is not seen to the northern end of 
the polarization disc in either the H or K bands.  The similarity in 
the degree and  position angle of the dichroic aspect of the 
polarization for YLW16A and YLW16B suggests that it is possible that 
there is a strong feature at this point that is responsible.

Figures 4(b) and 4(c) are images of the degree of polarization and the 
polarized flux in the H and K bands.  
In the degree of polarization maps there is a region of low 
polarization that marks the position of the polarization disc, the 
lowest levels of polarization are observed at the ends of the 
polarization disc.  Higher polarization regions are observed to either 
side of the polarization disc.
 
In the H band polarized flux map there are two lobes that extend 
roughly northwest and southeast. The southern lobe is the largest. 
The polarized light is brightest in a central region that is roughly 
centred on the core.  The polarized light is pinched to the south of 
the core of YLW16A; a similar pinching is not seen to the north.  In 
the K band the structure revealed gives the polarized light the 
appearance of being an inverted V.  The brightest region of polarized 
light is offset from the core, lying almost central to the eastern 
lobe.  A second less brilliant region is seen in the western lobe. 
 
YLW16A has the lowest levels of maximum polarization of all the 
extended sources, but it does not have the lowest core polarization. 
In Table 2 the core polarization of YLW16A in the K band was shown to 
be P$_{core}$ $\sim$ 10\%, with a maximum polarization of P$_{max}$ 
$\sim$ 21\%.  The polarization data does not cover the full extent of 
the YLW16A, so this may well be an underestimate for $\it{P}_{max}$. 
 
\subsection{WLY54} 
 
The structure revealed in Figure 1c indicates that WLY54 is a cometary 
nebula.  The tail of nebulosity extends roughly to the east of the 
core; the H band data indicates that the tail extends for at least 
1500 AU.  In the H and K band total flux distribution maps there is a 
second `object' (labelled S1) approximately 720 AU to the northwest of 
the core, the nature of this object is not known.  There does not 
appear to be any nebulosity extending from WLY54 to surround S1. 
However, in the K band there is a second knot of nebulosity 
between WLY54 and S1. 
 
\setcounter{figure}{3} 
\begin{figure} 
  \vspace{-2cm}\hspace{-1.3cm}\includegraphics[scale=0.6,angle=0]{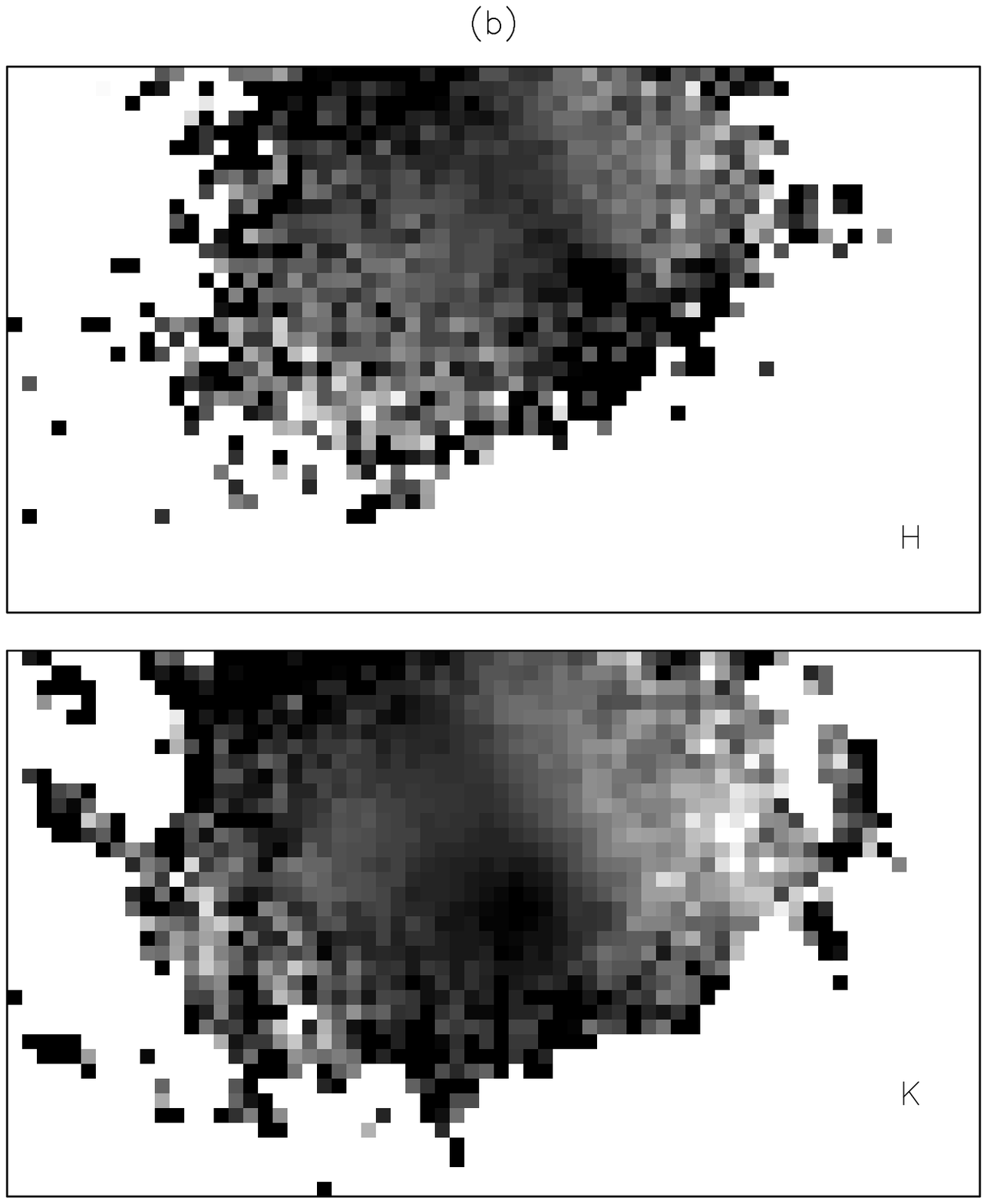}\vspace{-1.8cm} 
\caption{(b). Degree of polarization maps for YLW16A.} 
\end{figure} 
\setcounter{figure}{3}
\begin{figure} 
  \vspace{-2cm}\hspace{-1.5cm}\includegraphics[scale=0.6,angle=0]{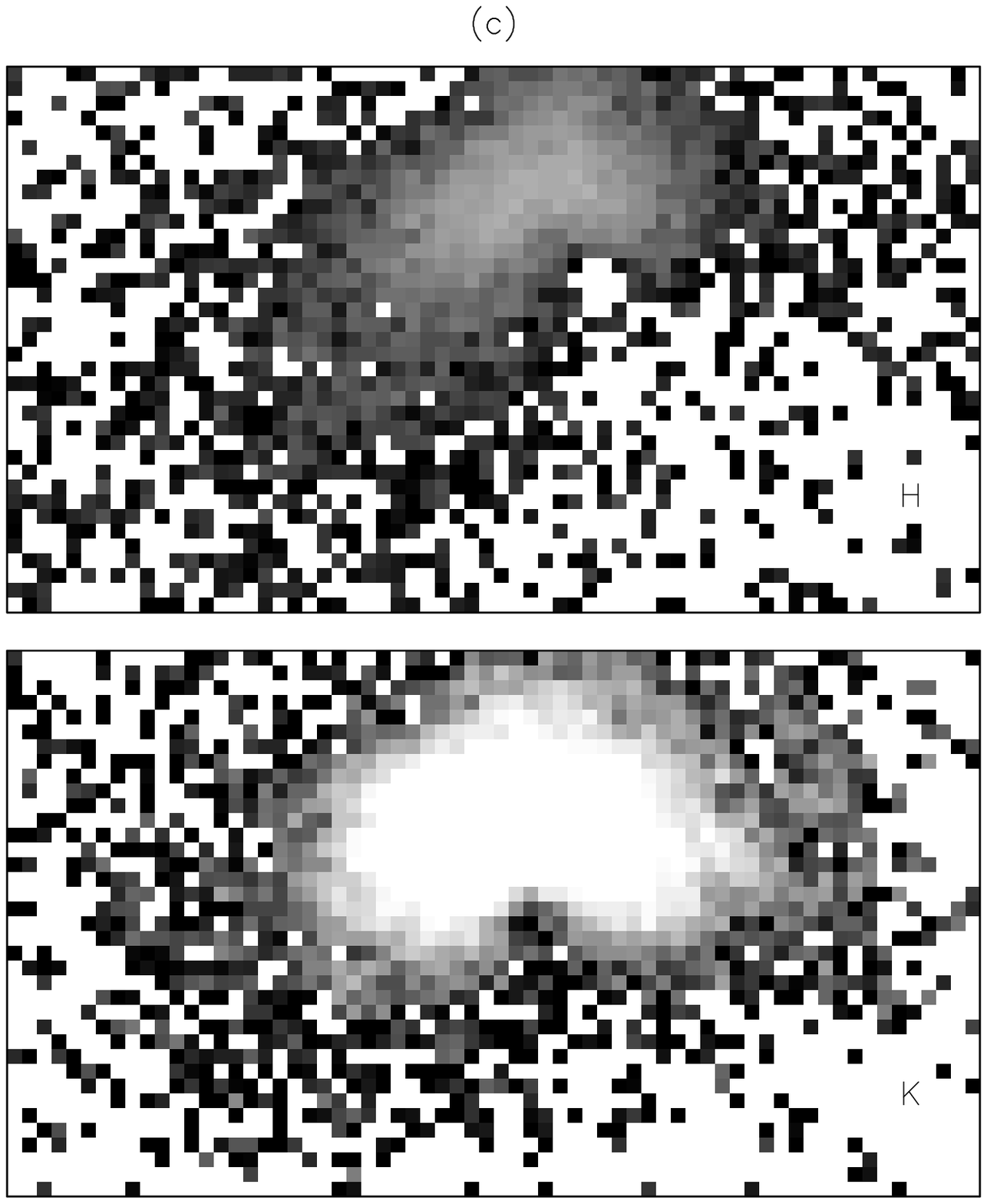}\vspace{-1.8cm} 
\caption{(c). Maps of the polarized flux for YLW16A.} 
\end{figure}  
 
The polarization vector maps in the J, H and K band are presented in 
figure 5(a).  The vector pattern is centrosymmetric.  There is 
evidence in the J and H band maps of a narrow polarization disc with a 
position angle $\theta$ $\sim$ 0$^{\circ}$, which lies over the 
central flux peak.  The polarization discs observed for other objects 
appear to be perpendicular to the orientation of the extended 
nebulosity; this is not the case for WLY54.  In WLY54 the angle 
between the polarization disc and the direction of the extension is 
approximately 109$^{\circ}$. 
 
\begin{figure} 
 \hspace{-1.8cm}\includegraphics[scale=0.58]{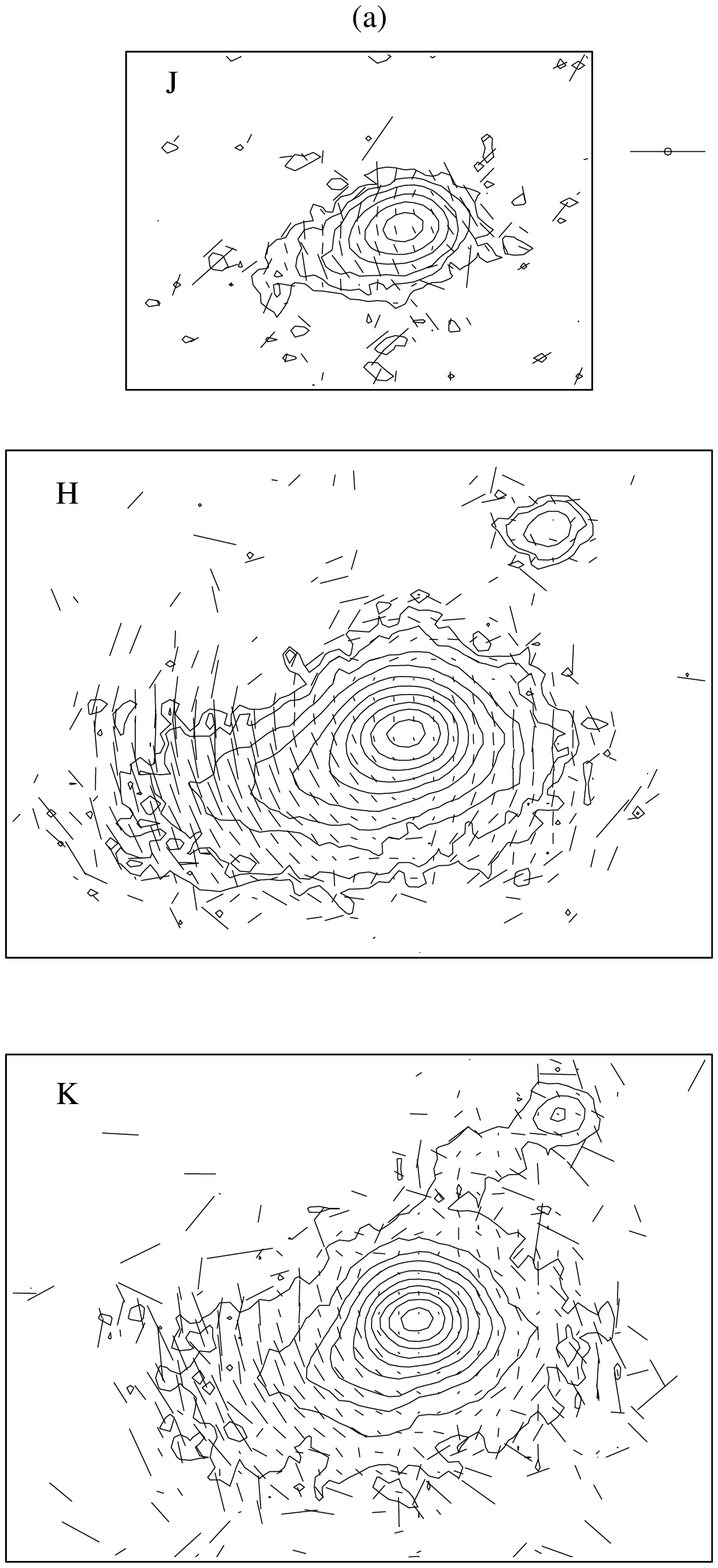} 
    \caption{WLY54 polarization. (a) J, H and K band contour plots of 
the flux distributions with polarization vectors overlaid, the key is 
the same length as a 100\% vector, (b) Degree of Polarization maps, 
and (c) Polarized flux distribution maps, for WLY54.   } 
\end{figure}  
 
The polarization vector pattern indicates that WLY54 might be a 
bipolar system that is at an inclination angle that minimizes the 
amount of counterlobe visible.  The position angle of the polarization 
disc relative to the direction of the extension would make sense if 
there were a source of foreground extinction that is obscuring the 
northern regions of WLY54. 
 
The core polarization levels seen are approximately 2\%, 5\%, and 9\% 
at the K, H, and J wavebands respectively.  This is at the lower end 
of the levels observed for Class I objects, and is much lower than the 
levels for the other extended sources.  The maximum polarizations for 
WLY54 are assessed over the cometary tail at a distance of 
approximately 6 arcseconds (1000 AU) from the flux peak.  The maximum 
polarizations at H and K are approximately 30\% and 43\% respectively. 
 
The degree of polarization maps in the H and K bands (see figures 5(b) 
and (c)) reveal a low polarization region over the core of WLY54.  A 
higher polarization region is seen over the cometary extension, and a 
smaller higher polarization region is seen to the west of the core. 
The knot S1 is visible in the H band degree of polarization map; in 
the K band map S1 and S2 appear as a single ``stream''.  In figures 
5(b) and (c) the polarized flux maps at the H and K bands are 
presented.  The structure of the polarized light is similar to the 
total flux distribution.  The structure of the knots is not seen 
clearly in either the H or K band polarized light map.  In the H band 
there is a slight bipolar pinching to the south of the core; the 
pinching is not seen in the K band. 
 
\subsection{WLY67} 
 
WLY67 has a cometary morphology, see Figure 1d and Figure 6. The 
cometary tail extends to the north of the core for at least 7 
arcseconds (1120 AU).  In the K band the structure of the tail appears 
to narrow with distance from the core, a similar narrowing is seen in 
the H band data.  To the south of the core in the K band there is a 
small broad horn of nebulosity that is not visible in the H band. 
 
The polarization vector map presented in figure 6 reveals that the 
vectors over the cometary extension have a centrosymmetric 
orientation.  The vectors to the south of the core also appear to be 
centrosymmetric.  In the H band there is evidence of a narrow 
polarization disc with a position angle of $\theta$ $\sim$ 
90$^{\circ}$ over the core. The polarization disc is not seen in the 
K band. 
  
In polarized light the structure revealed for WLY67 is similar to that 
observed for the total flux distribution.  There is no significant 
pinching towards the ends of the polarization disc defined by the 
vector map. 
 
\subsection{VSSG18} 
 
VSSG18 is the only Class II source in the sample that is associated 
with a highly extended nebula (see Figure 1e, Figure 7).  All other 
non-Class I sources appear to be point-like in nature; suggesting that 
any associated extended nebulosity has already dispersed leaving only 
a more compact structure.  This is consistent with star formation 
models, which predict that by the Class II stage of evolution an 
object will have lost its extended envelope. The extensive nebulosity 
of VSSG18 suggests that it's evolutionary status is actually similar to 
that usually associated with Class I sources. If this is so, the weak  
mid-IR and far-IR emission could be explained if the disc axis is 
oriented close to pole-on (see Whitney et al.2003) but we would then  
expect the central protostar to appear much brighter relative to 
the surrounding nebula. A possible alternative explanation is a gap 
in the inner accretion disc, perhaps due to binarity, which would 
reduce the amount of warm dust emitting at the IRAS wavelengths. 
 
The nebulosity associated with VSSG18 has a cometary morphology, which 
extends roughly to the northwest.  The nebulosity broadens with 
distance from the core, giving the material a fan-like appearance. 
The H and K band flux distribution maps show that the cometary 
nebulosity extends for at least 2500 AU.  In the J band the structure 
of the nebulosity differs from that observed in the H and K bands. 
The nebulosity appears to extend to the north; approximately 320AU 
from the core the nebulosity is ``pinched'', before broadening.  HST 
NICMOS data reveals a large sigma-shaped nebula; the nebulosity 
extends for at least 3800 AU, the exact location of the main 
illuminating source is not identified (Allen et al. 2002). 
 
The polarization vector maps in the H and K bands are presented in 
figure 7.  The vectors over the extended nebulosity are 
centrosymmetric.  Atypically, the vectors over the core are aligned 
with a position angle of $\theta$ $\sim$ 145$^{\circ}$, placing them 
parallel to the direction of the extension.  The other objects that 
have a centrosymmetric polarization vector structure show aligned 
vectors that are approximately perpendicular to the direction of the 
extended nebulosity.  To the southern end of the aligned vectors there 
is no clear evidence of a return to the centrosymmetric pattern.  The 
amount of information in the J band polarization vector map (not 
shown) is limited, but it can be seen that vectors over the core are 
at a position angle of 0$^{\circ}$ and the vectors over the extended 
nebulosity are centrosymmetric. 
 
The structure of the polarized light reveals a bright region 
approximately centred over the core surrounded by a faint region that 
marks the extended nebulosity.   There is no evidence of the typical 
bipolar pinching at the ends of the polarization disc. 
 
\subsection{EL29} 
 
EL29 (see Figure 1g, Figure 8) is one of the four objects that were 
imaged using the shift and add image sharpening technique during the 
UKIRT 1999 observing run.  Although EL29 has been studied before at 
near-IR wavebands this is the first time it has been looked at using 
high resolution linear polarimetry.  The structure of the total flux 
distribution revealed indicates that EL29 is a point-like object. 
Previous studies have indicated that El29 may be associated with 
extended nebulosity (Elias 1978; Duchene et al. 2004). 
 
EL29 differs from the other point-like objects discussed previously 
(see $\S$4.1); instead of the aligned or random polarization vector 
pattern, the vectors are centrosymmetric and they extend for 
approximately 7 arcseconds (1120 AU) from the core.  There is a narrow 
polarization disc over the core, with a position angle $\theta_K$ 
$\sim$ 20$^{\circ}$.  To the northern end of the disc the level of 
polarization decreases to approximately $\it{P}$ $\sim$ 0\%, a similar 
decrease is seen to the southern end of the disc.  The polarization 
vectors return to the centrosymmetric pattern after the 'nulls'. 
 
\setcounter{figure}{4} 
\begin{figure}
 \vspace{0cm}\hspace{-1.5cm}\includegraphics[scale=0.58]{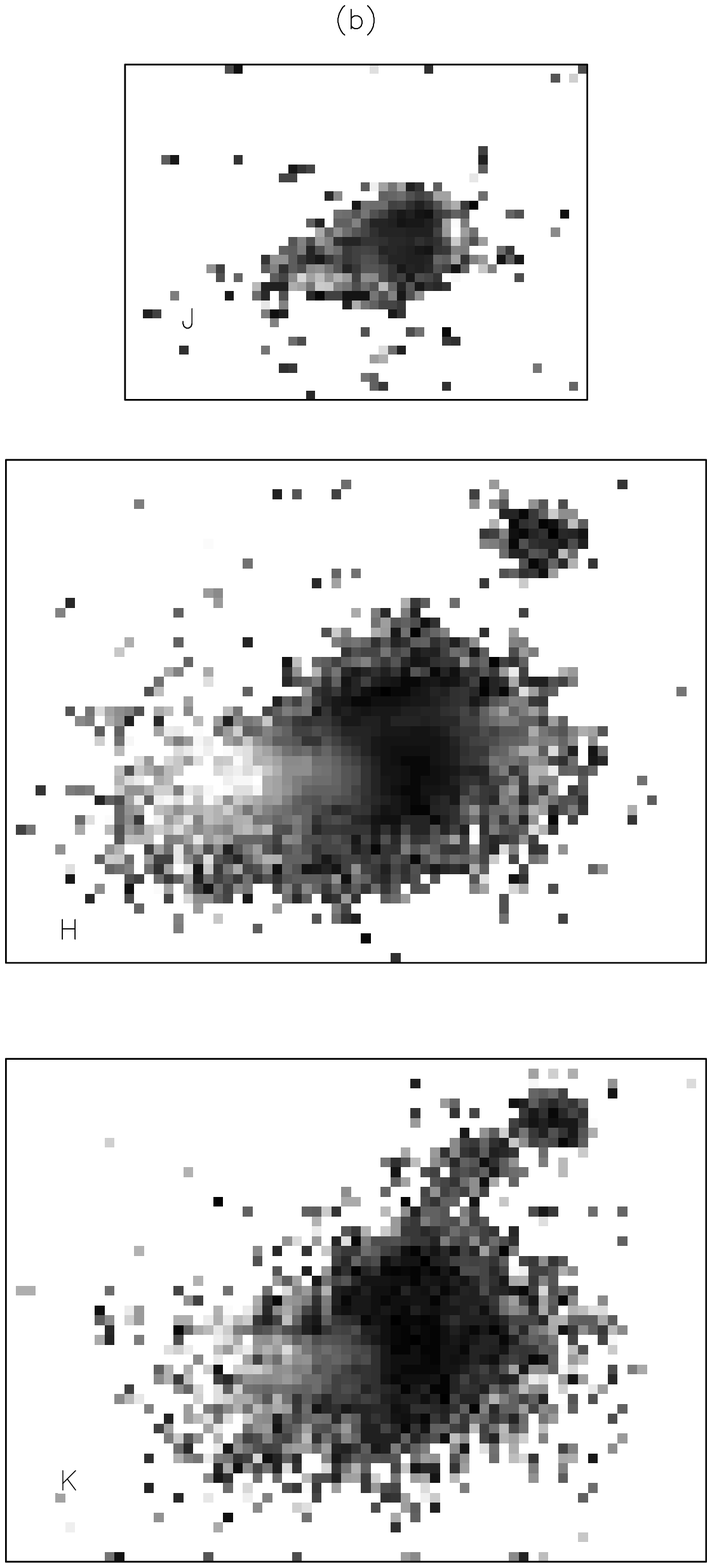} 
\caption{(b). Degree of polarization maps for WLY54.} 
\end{figure}  
\setcounter{figure}{4} 
\begin{figure} 
 \vspace{0cm}\hspace{-1.5cm}\includegraphics[scale=0.58]{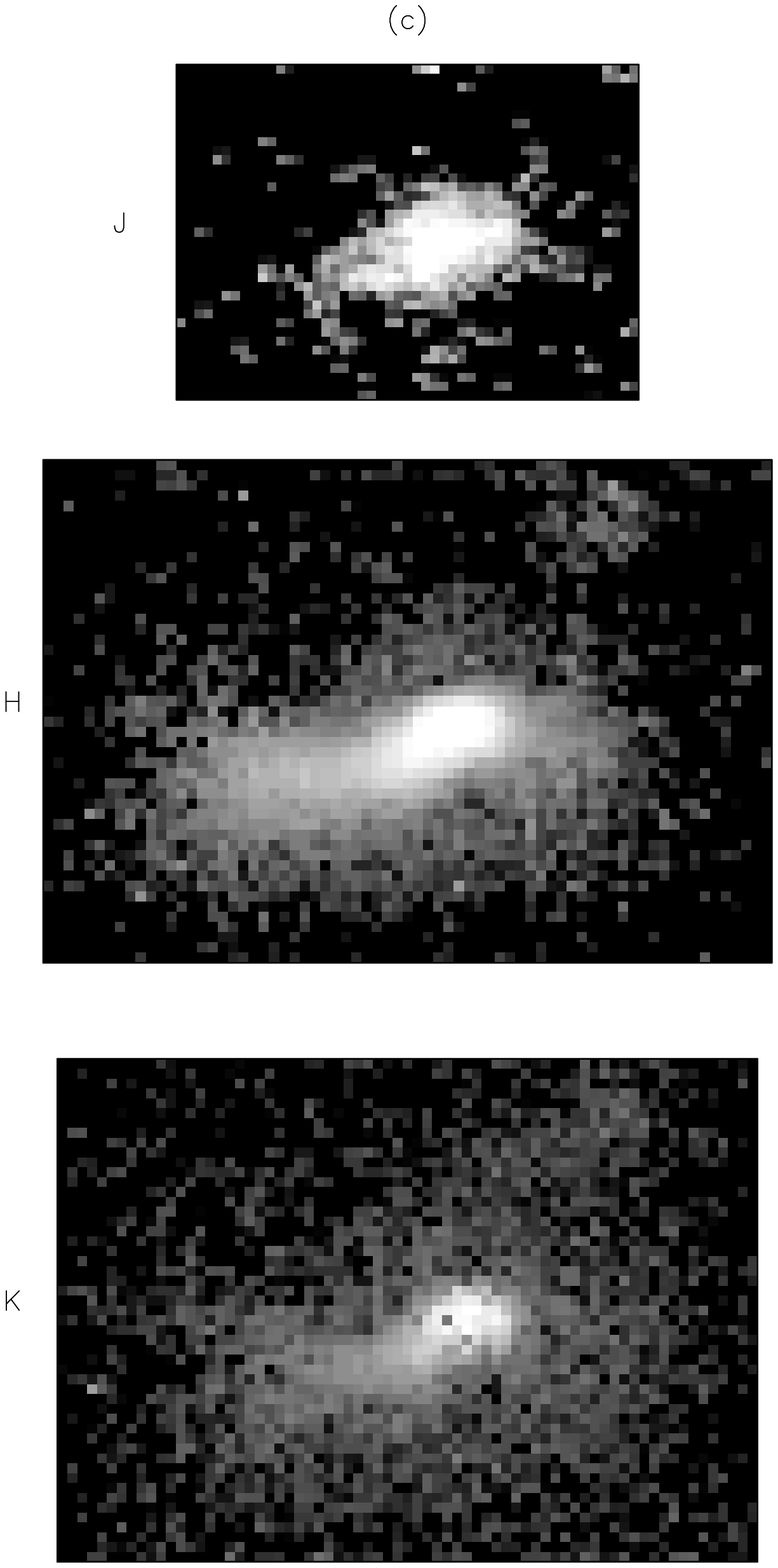} 
\caption{(c). Maps of the polarized flux for WLY54.} 
\end{figure}  

\begin{figure} 
\vspace{-4mm} 
\hspace{-1.3cm}\includegraphics[scale=0.38,angle=-90]{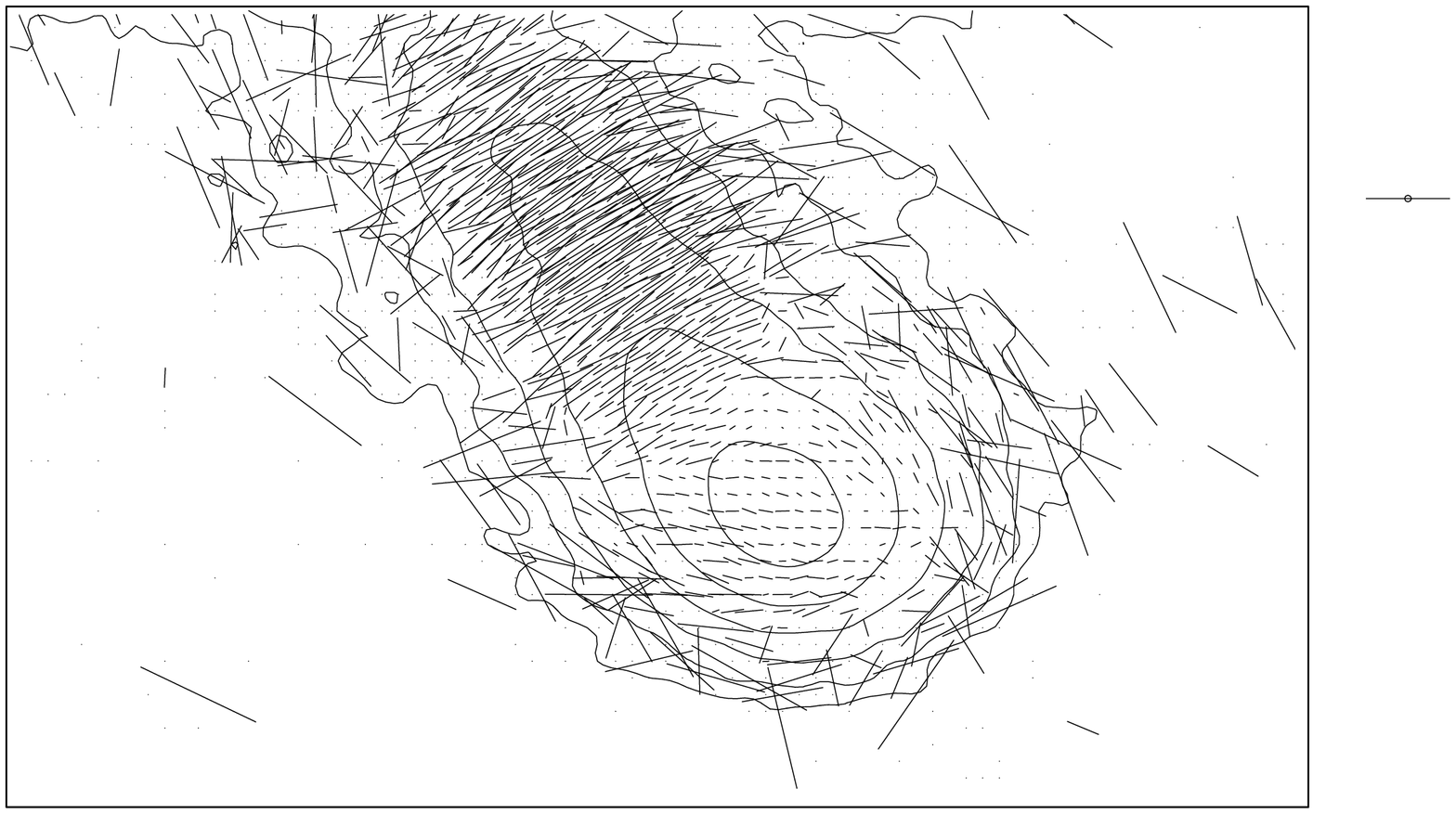}\vspace{-1.3cm} 
 \caption{WLY67 polarization (K band). A contour plot of the flux 
 distribution with polarization vectors is overlaid. The key 
 length is the same as a 50\% vector.} 
\end{figure}  

\begin{figure} 
 \vspace{-1.6cm} 
\hspace{-1.3cm}\includegraphics[scale=0.55]{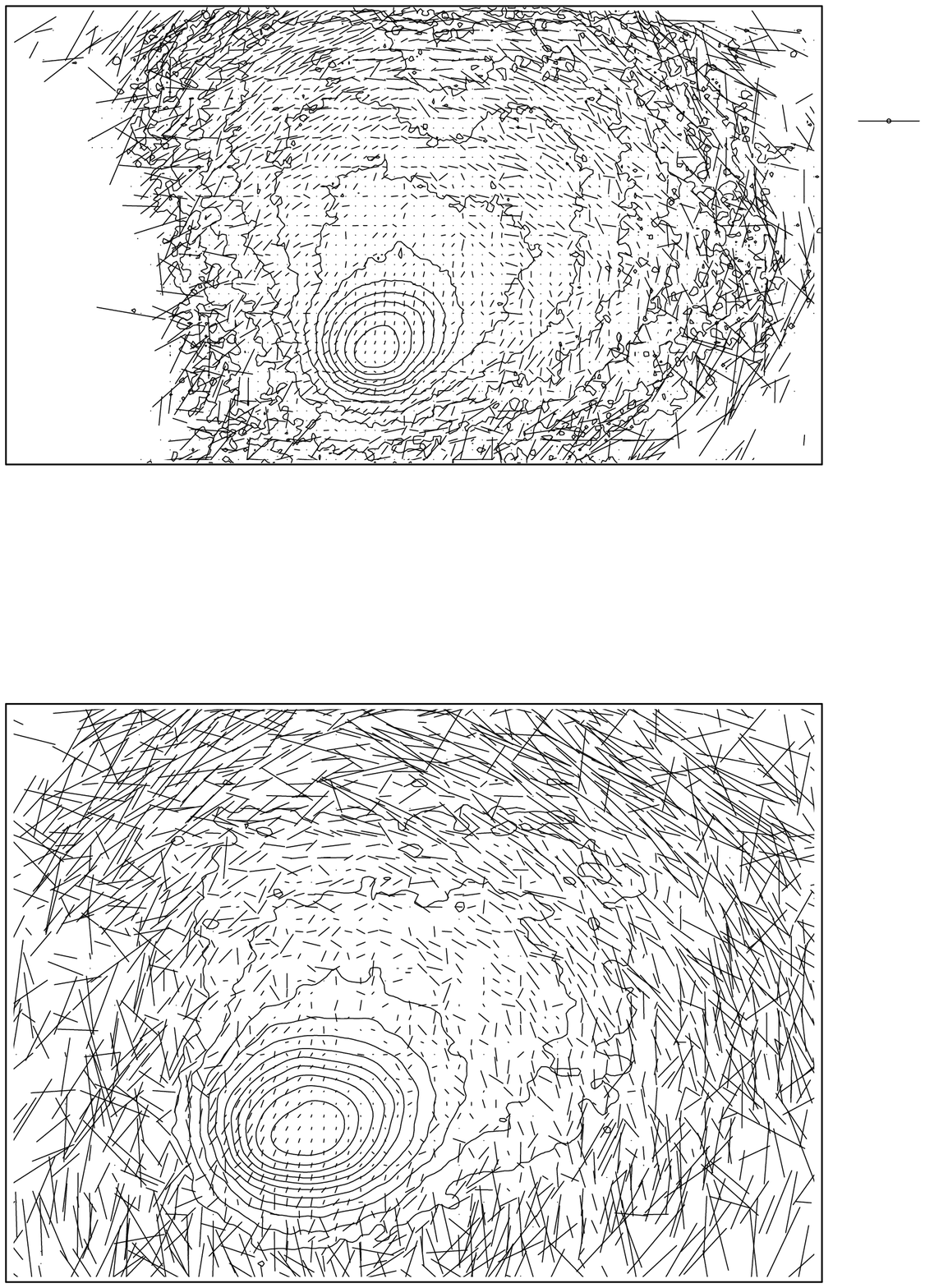}\vspace{-1cm} 
 \caption{VSSG18 polarization. Contour plots of the flux distributions 
 with polarization vectors overlaid for VSSG18.  The key is the same 
 length as a 50\% vector.} 
\end{figure}  
 
\begin{figure} 
 \hspace{-3cm}
\vspace{-28mm}\includegraphics[scale=0.75]{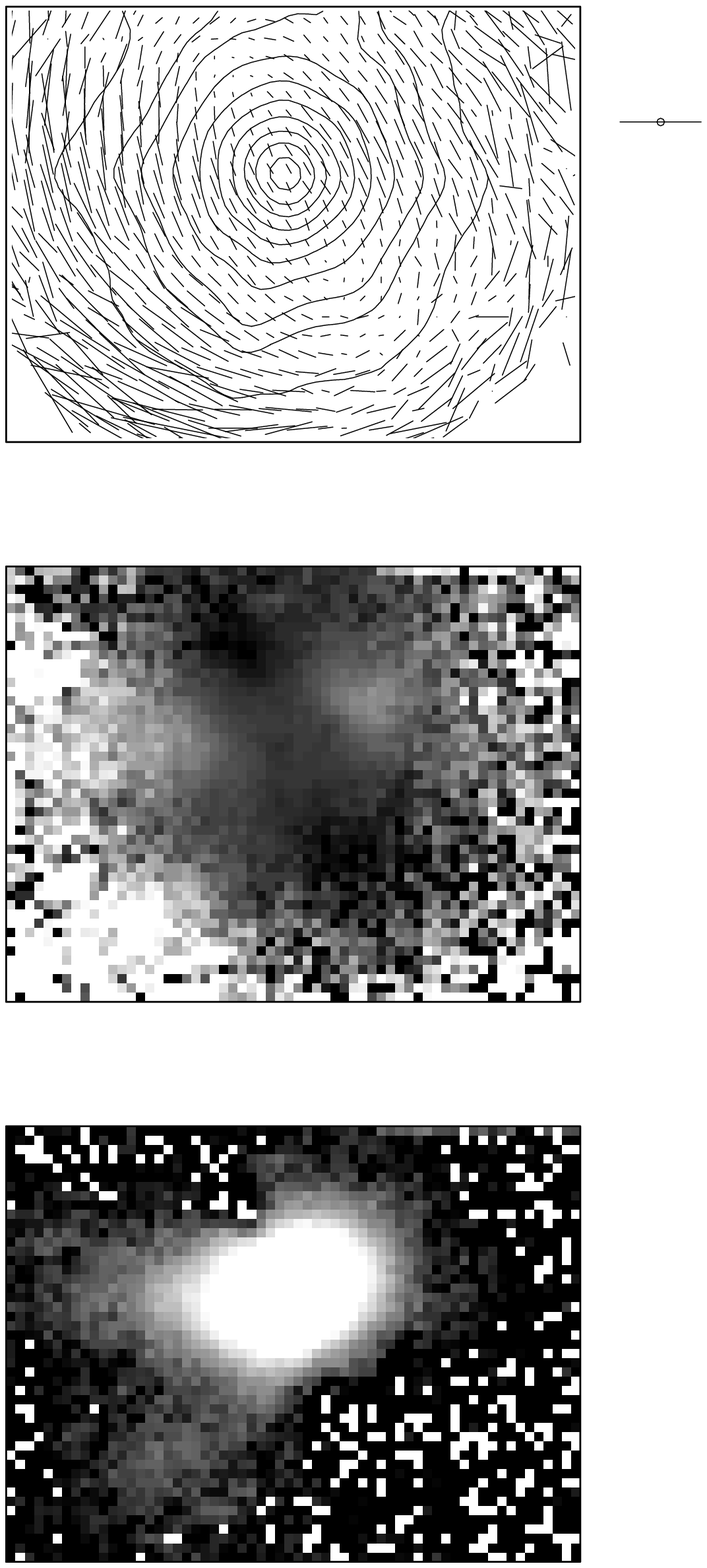} 
\vspace{14mm} \caption{EL29 polarization.(a) Contour plot of the flux distribution 
 with polarization vectors overlaid, the key is the same length as a 
 50\% vector, (b) Degree of Polarization map, scaled between 0\% 
 (black) to 40\% (white) and (c) Polarized flux distribution map for 
 EL29, plotted on a logarithmic scale between 0 (black) to 2.4 
 (white).} 
\end{figure}  
 
The pattern seen for the polarization vectors strongly suggests that a 
faint bipolar nebula surrounds EL29.  Figures 8(b) and (c) are maps of 
the degree of polarization and the polarized intensity.  In figure 
8(c) there is a disc-like region of lower polarization surrounded by a 
higher polarization region.  At the ends of the disc, two polarization 
nulls are seen.  The structure revealed in the polarized intensity map 
is not the point-like object seen in the total flux distribution. 
Instead in polarized light EL29 appears to be elongated, with a 
southeast-northwest orientation.  There is a slight `pinching' of the 
polarized light that is characteristic of a bipolar object. 
 
This result for the polarised flux image has recently been 
independently confirmed by the conference report of Hu\'{e}lamo et 
al.(2007), which contains higher resolution data from the 
VLT.  Deep infrared imaging by Ybarra et al.(2006) also reveals the 
extended  nebulosity and shows that there are H$_2$ emission features 
within the outflow cavities, which are oriented southeast-northwest as 
our polarised flux image implied. 
 
\subsection{Discussion} 
 
\subsubsection{Wavelength Dependence} 
 
\begin{figure} 
\hspace{-4mm} \includegraphics[scale=0.45]{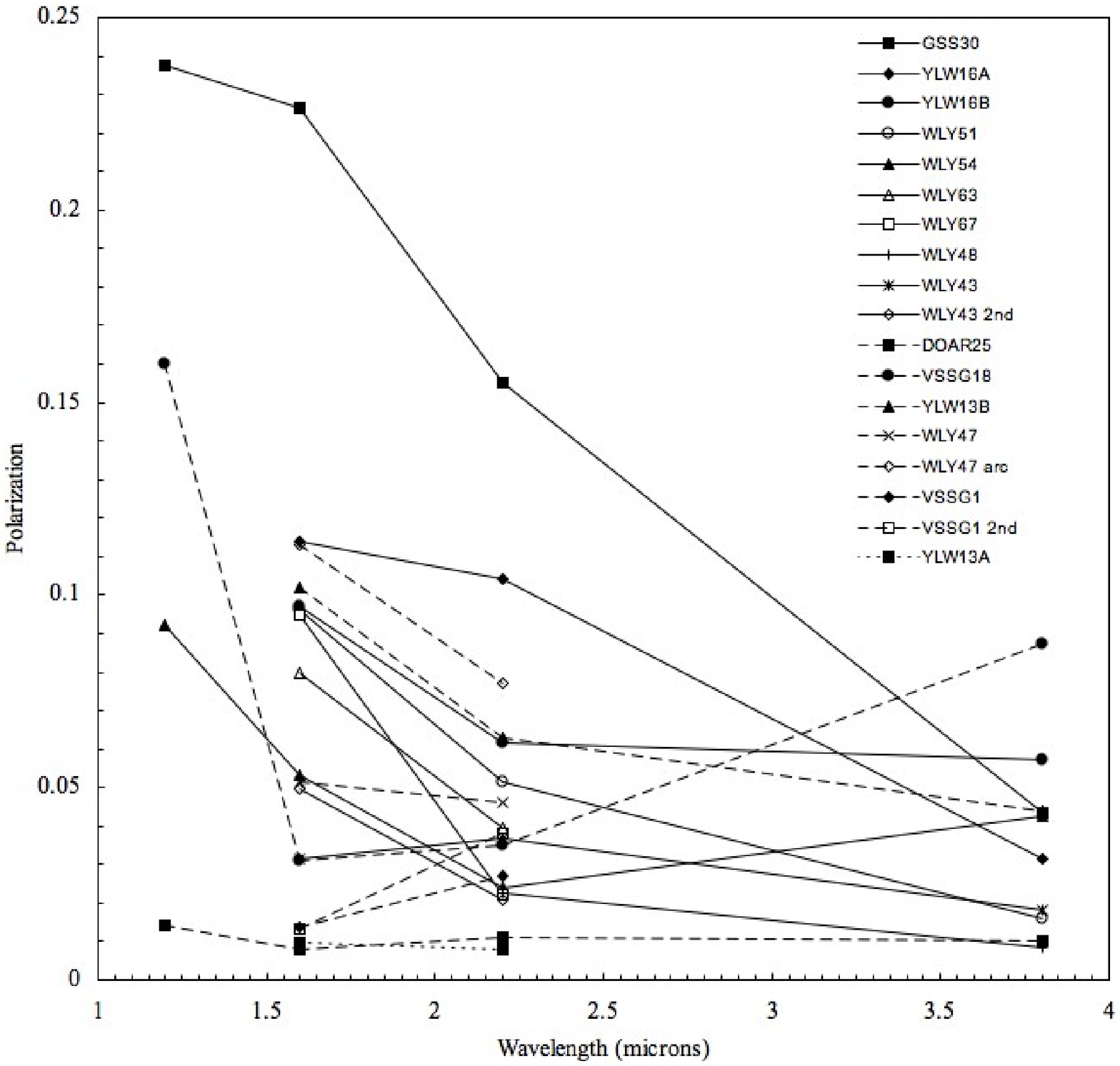} 
 \caption{Wavelength dependence of the core polarizations.  
Solid lines are class I YSOs, dashed lines are class II and
dotted lines class III.
The behaviour of the polarization with wavelength indicates that for the
majority of the sources dichroism is important in the core regions.} 
\end{figure}  
 
The wavelength dependence can reveal information about the nature of 
the mechanisms responsible for the scattering and absorption of light 
in an YSO.  The wavelength dependence in the core regions for the 18 
objects that were observed at multiple wavebands is shown in figure 9.
The key indicates the identities of individual sources. The wavelength 
dependence seems to be independent of the IR classification; similar
behaviour is observed for Class I, II and III YSOs. Most of the objects show 
a change of a few percent, typically 3 -- 4\%, between 1.6 $\mu$m and 2.2 $\mu$m. 
 
In the majority of the objects the degree of polarization decreases 
with increasing wavelength.  This 
is the expected wavalength dependence if dichroism is the main mechanism  
responsible for the polarization of light. However it is also possible 
for scattering to produce the same wavelength dependence if: (i)   
the polarization within the adopted aperture integrates  
over a complicated polarization structure (Whitney et al.1997); or 
(ii) the measurement is of a polarization disc of a bipolar source where
most photons have been multiply scattered (Lucas \& Roche 1998). 
We think it very likely that the observed wavelength dependence in the point
sources is due to dichroic extinction in most cases. The alternative 
explanation, that the measured polarisation is due to scattering in
spatially unresolved reflection nebulae with complicated polarization 
structures, is not likely to be such a common occurrence. 
 
Four of the YSOs investigated show an increase 
in the level of polarization with increasing wavelength: VSSG1, BKLT 
J162618-242818, WLY43, and the WLY47 `arc'.  This indicates that 
scattering is important in these sources, since a rising or flat polarization 
with increasing wavelength is predicted by the Mie theory for single  
scattering by sub-micron grains with a range of sizes. In VSSG18 there is  
a decrease in the polarization between 1.2 $\mu$m and 2.2 $\mu$m, between  
2.2 $\mu$m and 3.8 $\mu$m the degree of polarization increases.  The same behaviour 
is observed for WLY54.  Three of the objects show no significant 
change in polarization with wavelength.  The first is the Class III 
object YLW13A, the second is the Class II object DOAR25, and the third 
is the Class II object WLY47.  The polarization levels for both YLW13A 
and DOAR25 are very low, $\it{P}$  $\sim$ 1\%, whereas the degree of 
polarization determined for WLY47 is $\it{P}$ $\sim$ 5\%.  The Class I 
object YLW16A shows only a one percent change in polarization between 
1.6 $\mu$m and 2.2 $\mu$m, $\it{P}$$_H$ $\sim$ 11\% and $\it{P}$$_K$ 
$\sim$ 10\%.  However, the level of polarization is shown to 
significantly decrease between 2.2 $\mu$m and 3.8 $\mu$m, 
$\it{P}$$_{Lp}$ $\sim$ 3\%.  The same trend is seen for the Class I 
object GSS30 between 1.2 $\mu$m and 1.6 $\mu$m; only a one percent 
change in polarization is observed, with larger decreases seen after 
1.6 $\mu$m. 
 
\begin{figure} 
\hspace{-4mm} \includegraphics[scale=0.45]{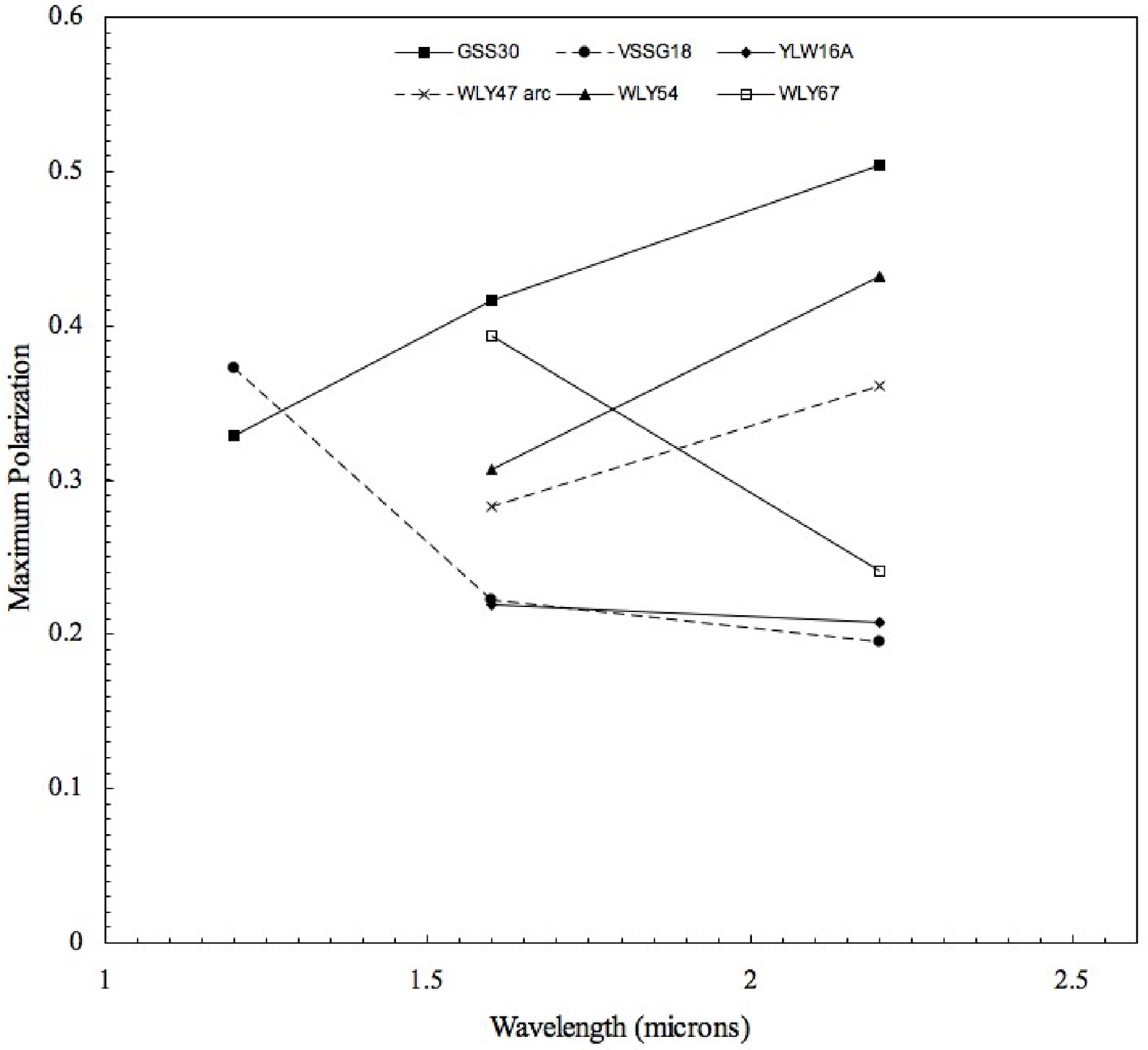} 
 \caption{Wavelength dependence of the maximum polarizations, for 
 spatially resolved sources. Class I and Class II YSOs are represented 
 with solid and dashed lines, respectively.  The behaviour of the 
 maximum polarization suggests that in the outer regions of the YSOs 
 scattering is important.} 
\end{figure}  
 
Figure 10 shows the wavelength dependence for the maximum 
polarizations for the 6 objects with extended nebulae.  The degree of 
polarization is seen to increase with increasing wavelength for three 
of the sources. One of the YSOs shows little change in polarization 
with wavelength; the remaining two objects display a decrease in 
polarization with wavelength.  Of the three sources showing a positive 
slope two are Class I and extended, the third is the extended arc of 
nebulosity that is located close to the Class II object WLY47.  WLY47 
is typically given a Class II designation, however the spectral index 
$\alpha$ = 0.17 puts WLY47 in the transition object regime and the 
aperture is across the arc of nebulosity.  The first object that shows 
a definite negative slope is the Class II object VSSG18.  The other 
object that has a negative slope is WLY67; it should be noted that the 
polarization data in the K band is poor quality.  YLW16A shows only a 
1\% decrease in polarization.  This could be explained as a 
result of the proximity of the aperture centre to the polarization 
disc; the data for YLW16A only cover the inner region of the envelope. 
 
Figures 9 and 10 suggest that in the inner regions of YSOs dichroism 
is an important mechanism in the production of polarized light. However, 
we caution that this conclusion might be false if there are spatially 
unresolved scattered light structures with complicated polarization  
patterns. In the outer envelopes scattering is usually dominant.   
With the exception of the arc of nebulosity close to WLY47 all the sources 
that display core polarization levels above P$_{core}$ $\sim$ 4\% have 
negative slopes between 1.2 $\mu$m and 2.2 $\mu$m.  The sources that 
display negative slopes to their maximum polarizations have P$_{max}$ 
$<$ 30\%; aside from VSSG18 the apertures for determining the maximum 
polarization are centred no more than 4 arcseconds from the flux peak. 
 
\subsubsection{Evolutionary Indicator} 
 
\begin{figure} 
 \includegraphics[scale=0.45]{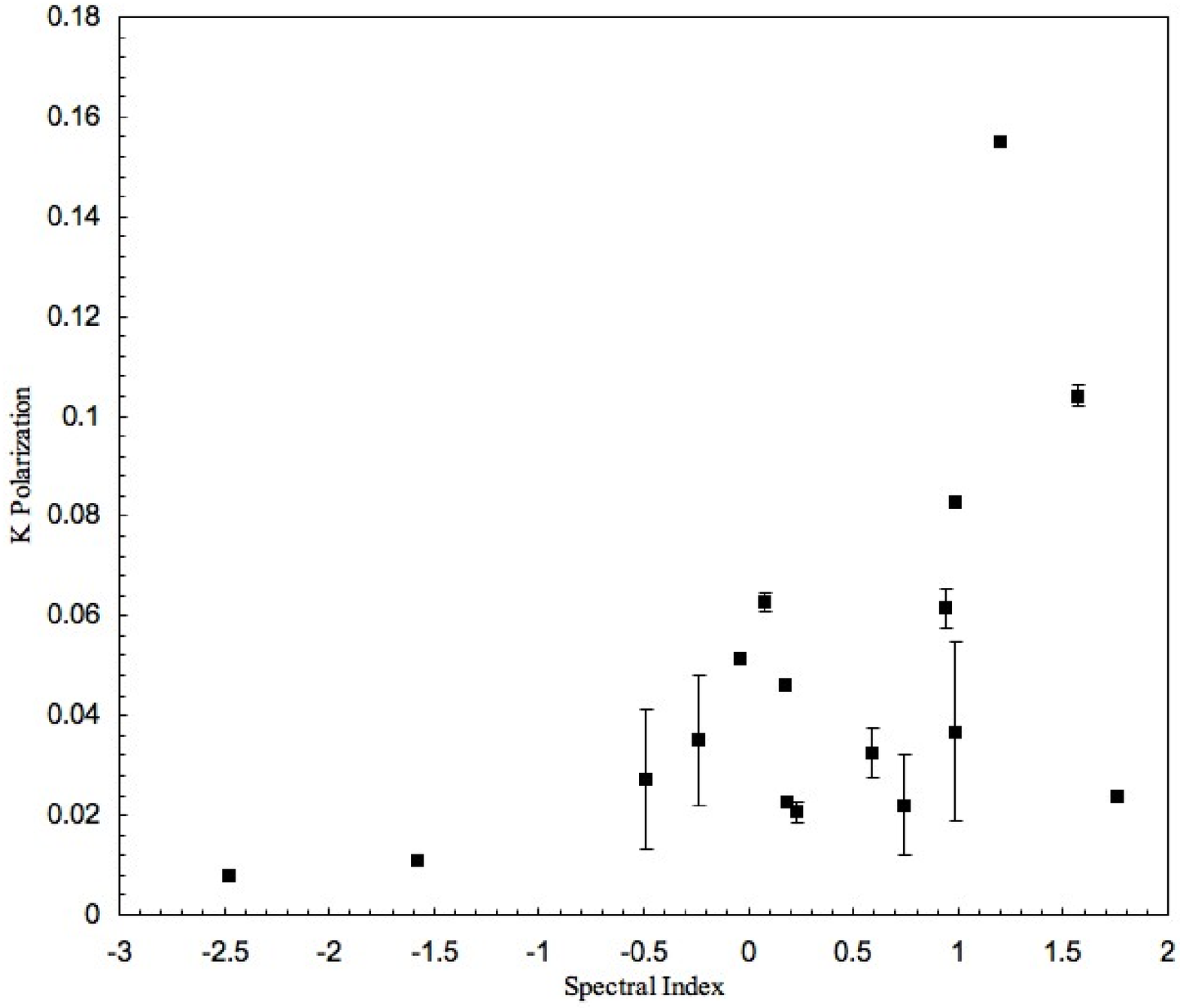} 
 \caption{A comparison of the degree of core polarization in the K 
 waveband for each object with its spectral index assessed in the 
 wavelength range 2 $\mu$m to 14 $\mu$m.  Error bars are shown for the 
 degree of K polarization; where no error bar is visible the error is 
 too small to be seen outside the point.  The higher the degree of 
 core polarization the more positive the spectral index.} 
\end{figure}

The current method used to determine the evolutionary stage of an YSO 
is based on a combination of the shape of its IR SED and its spectral 
index; both are dependent on the distribution of circumstellar 
material.  Generally, the degree of core polarization is seen to 
decrease as the age of the YSO, as indicated by the IR classification, 
increases.  The Class I sources typically display polarizations in the 
range 1\% $<$ $\it{P}$$_K$ $<$ 20\%, Class II sources 1\% $<$ 
$\it{P}$$_K$ $<$ 6\%, and the only Class III source has $\it{P}$$_K$ 
$<$ 1\%.  The wavelength dependence for the core polarizations in the unresolved
sources has  shown that the polarization very probably has a strong dichroic 
component.  The  lower polarizations and unresolved structure in the Class II sources 
probably indicate an evolutionary change to dichroic extinction as the 
dominant source of polarization, contrasting with scattering in the 
Class I sources.  The main difference in the dust properties of the 
Class I and Class II stages of evolution is thought to be only that 
the circumstellar envelope has dissipated, leaving only the disc. 
 
Again we caution that that Class designations are empirical and 
are influenced by the system inclination as well as the evolutionary 
status. Phyiscally, the spectral index depends as much on the optical 
depth in the line of sight to the protostar as the actual structure 
of the circumstellar matter. Systems viewed in the equatorial plane  
(i.e. with an edge-on disc) will display a more steeply rising SED, 
higher polarization and higher optical depth, than pole-on systems,  
which may lead to mis-classification. 
 
Figure 11 compares the degree of polarization in the K band for all 
the sources against their spectral indices, $\alpha_{2-14}$.  As 
discussed in $\S$1.2.1, the spectral index is the slope of the 
infrared spectral energy distribution (IR SED) and is commonly 
assessed in the wavelength range 2 $\mu$m to 10-25 $\mu$m.  The 
subscript (2-14) indicates that the spectral indices plotted are 
assessed in the wavelength range 2 $\mu$m to 14 $\mu$m, where the 
longer wavelength is based on the ISO 14 $\mu$m filter.  The degree of 
polarization tends to increase with increasing $\alpha_{2-14}$; this 
trend is also observed in the H and Lp wavebands and is suggested by 
the limited J band data available.  It is important to note that the 
majority of the sources in the sample have $\alpha_{2-14}$ $>$ 0, only 
four of the sources observed have $\alpha_{2-14}$ $<$ 0. 
 
\begin{figure} 
 
\hspace{-4mm}\includegraphics[scale=0.45]{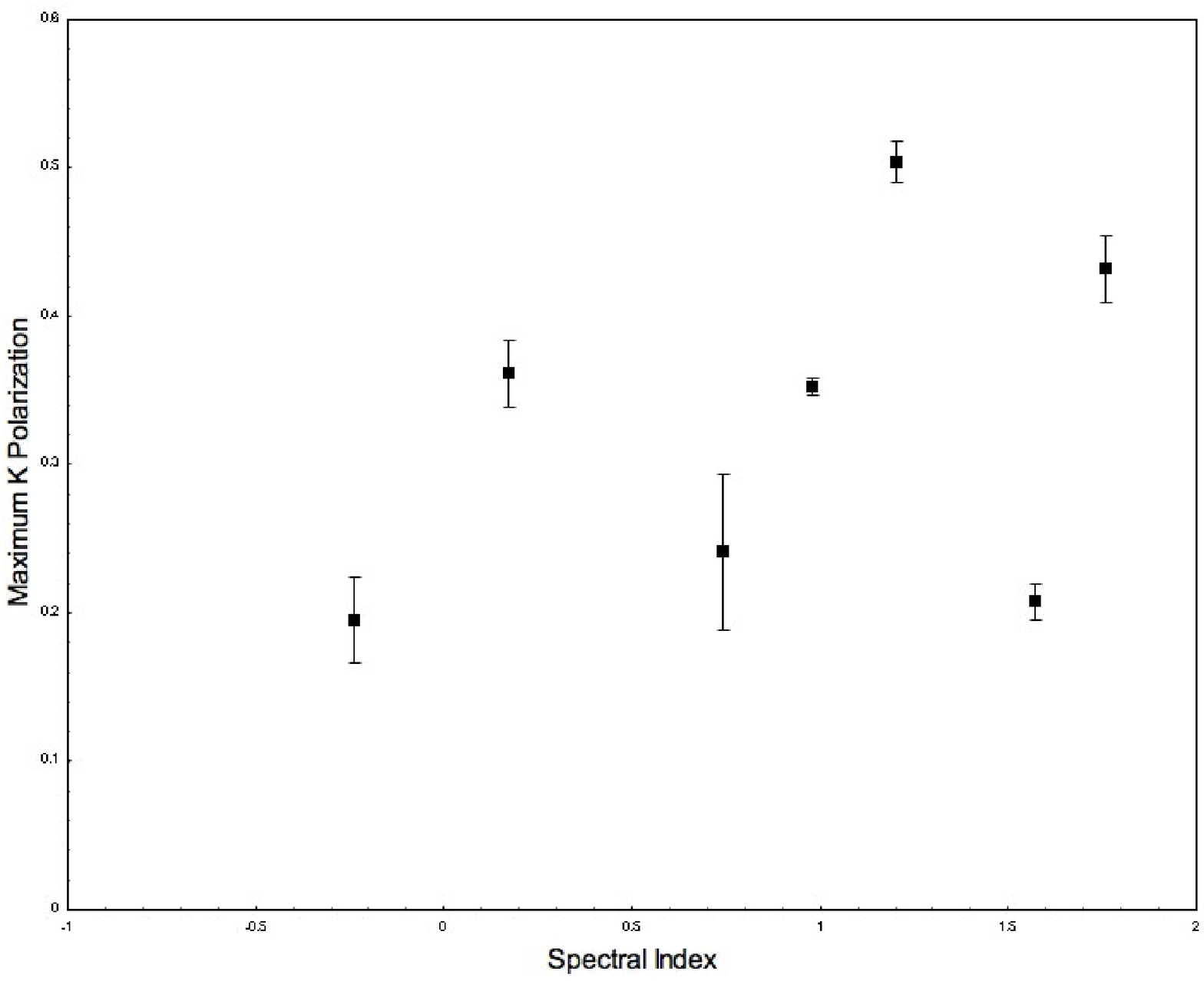} 
   \caption{A plot of the maximum degree of polarization in the K 
   waveband for each object with extended nebulosity against the 
   spectral index assessed in the wavelength range 2 $\mu$m to 14 
   $\mu$m.} 
\end{figure} 
 
The maximum degree of polarizations observed for each YSO in the K 
band is compared to their $\alpha_{2-14}$ in figure 12.  The maximum 
polarization of an YSO with extended nebulosity does not appear to be 
dependent on its IR evolutionary stage.  The same is also seen with 
the H band polarization.  $\alpha_{2-14}$ maps the warm gas component, 
whereas the extended nebulosity represents the cold gas component. 
 
Figures 11 and 12 show that the degree of polarization cannot be used 
to independently determine the evolutionary stage of an object, but it 
can be used to provide an approximate guide.  The older objects 
typically display lower core and maximum polarizations. Polarizations 
of P$>$6\% are only seen for Class I YSOs. 
 
\subsubsection{Polarization as a Function of Colour} 
 
\begin{figure} 
 \hspace{-4mm}\includegraphics[scale=0.45]{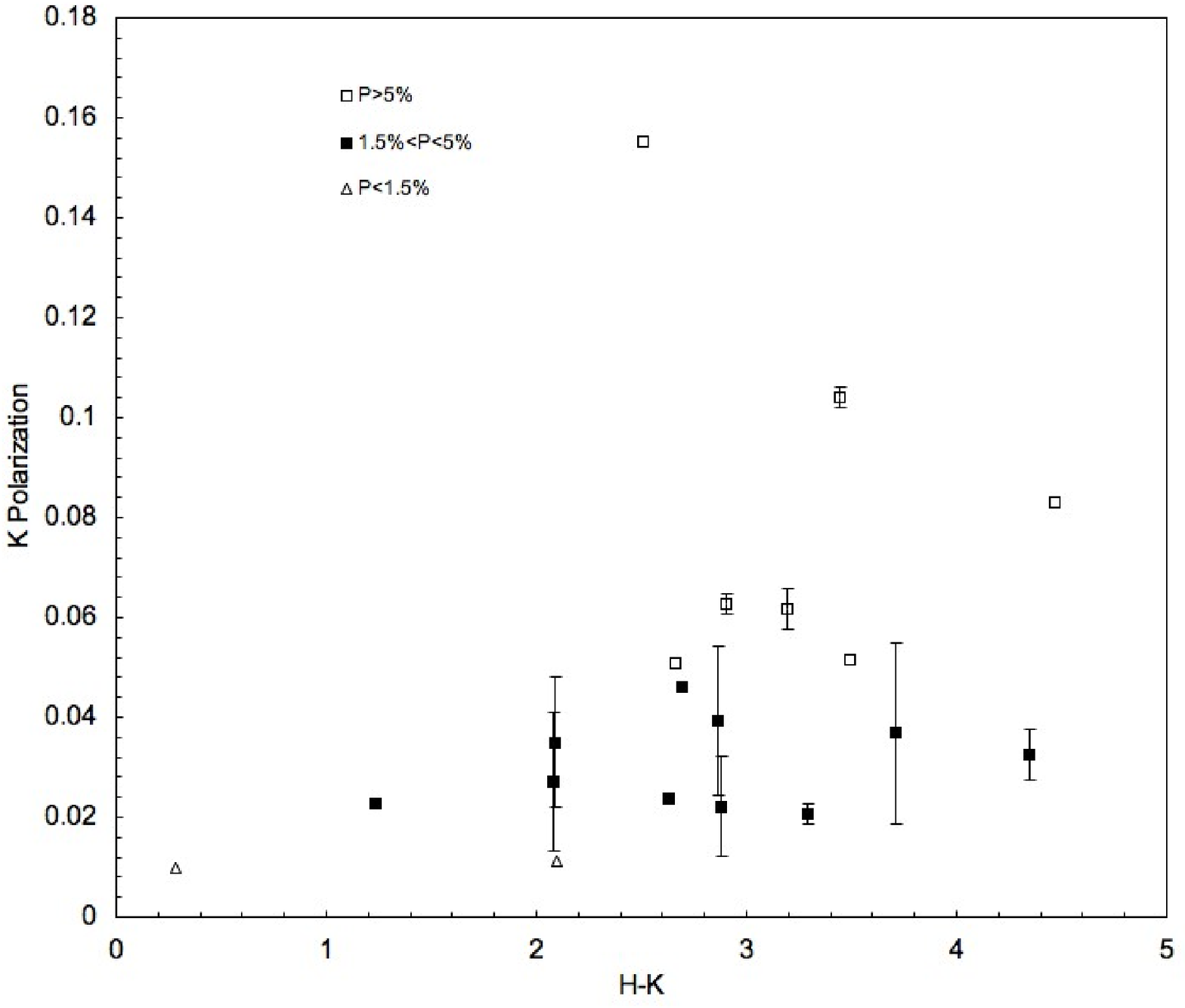} 
   \caption{A comparison of the degree of core polarization of each 
   object in the K waveband with its H-K colour.  The open triangles 
   indicate sources with P$_K$ $<$ 1.5\%, filled squares indicate 
   sources with 1.5\% $>$ P$_K$ $>$ 5\%, and open squares indicate 
   sources with P$_K$ $>$ 5\%.  Error bars are shown for the degree of 
   K polarization; where no error bar is visible the error is too 
   small to be seen outside the point.  The lowest core polarizations 
   are observed for the sources with lower H-K colours, and the 
   highest core polarizations are observed for the sources with the 
   largest H-K colours.} 
\end{figure} 
\begin{figure} 
 \hspace{0mm}\includegraphics[scale=0.43]{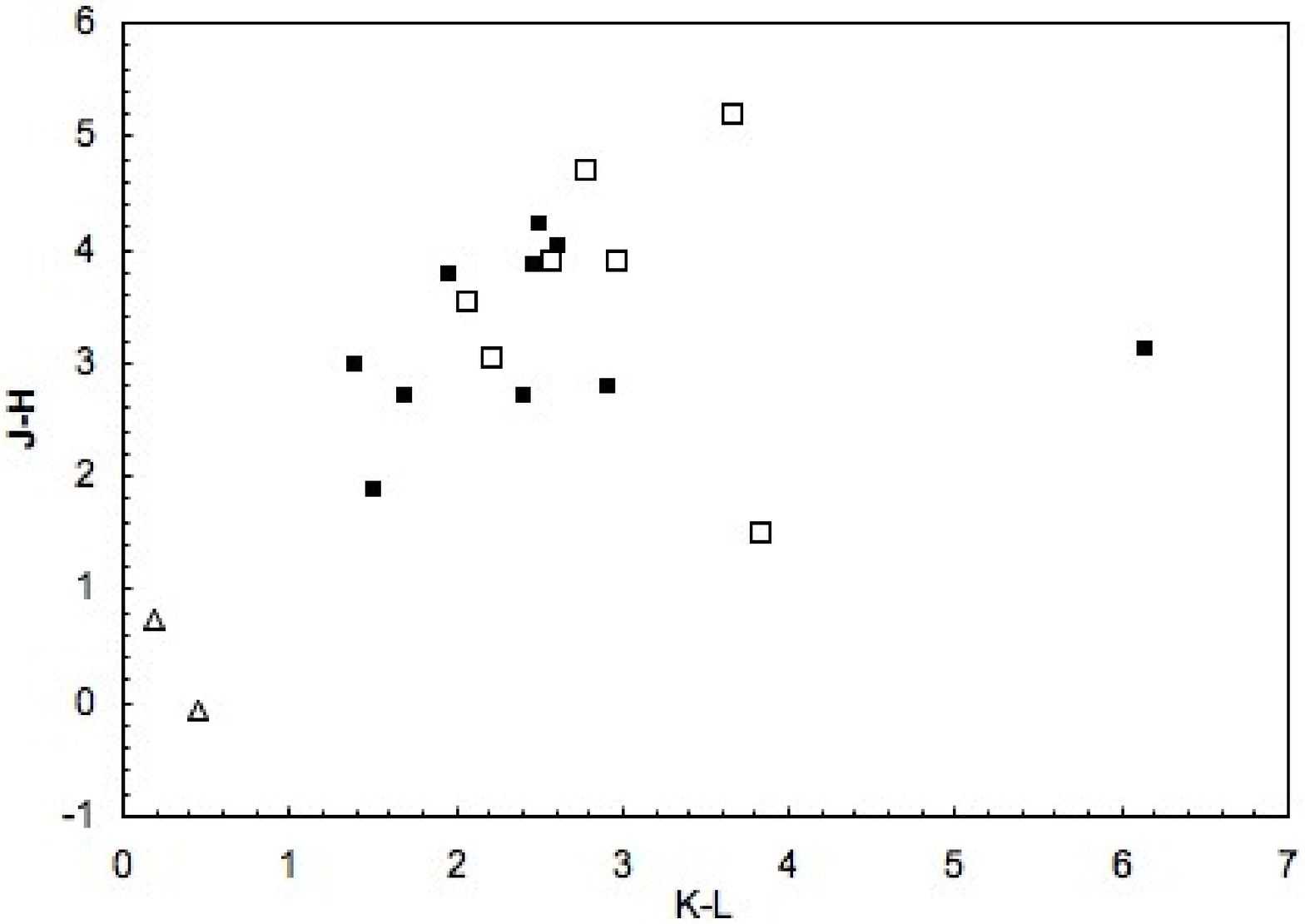} 
 \hspace{-2cm}\includegraphics[scale=0.45]{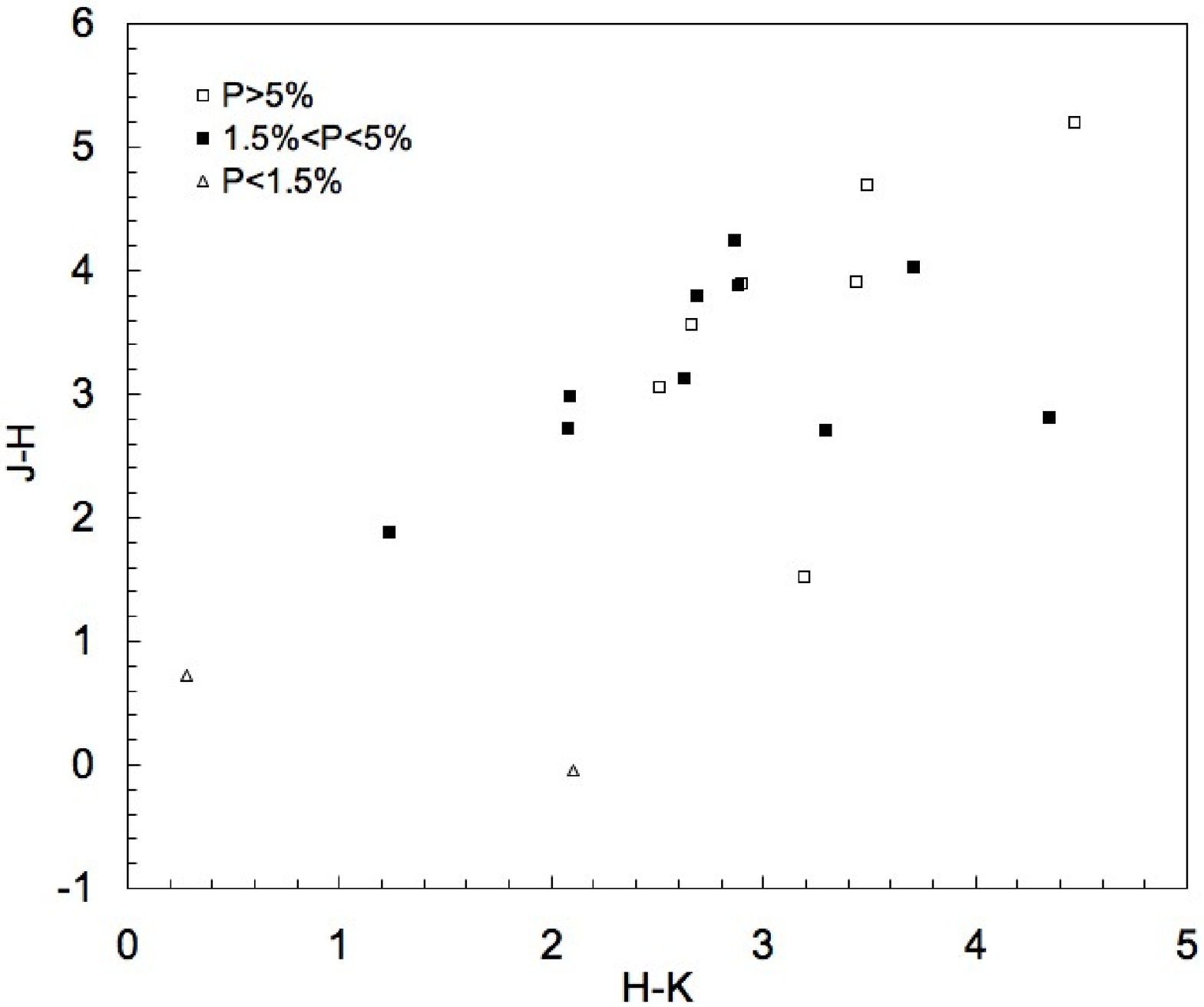} 
   \caption{Near-IR colour-colour diagrams with K core polarization 
   measurements.  The open triangles indicate sources with P$_K$ $<$ 
   1.5\%, filled squares indicate sources with 1.5\% $>$ P$_K$ $>$ 
   5\%, and open squares indicate sources with P$_K$ $>$ 5\%. } 
\end{figure} 
 
One of the advantages of conducting the survey in the $\rho$ Oph 
region is that complete near-IR J, H, K, and L spectral information is 
available for each source (see Wilking, Lada \& Young 1989; Greene et 
al. 1994; Barsony et al. 1997).  This enables comparisons between the 
polarimetric and spectral data.  In figure 13 the degree of 
polarization in the K band for each source is compared with its H-K 
colour.  The H-K colour is a measure of the redness of the source. 
The redness of an YSO decreases as it evolves.  Therefore, the H-K 
colour is an indicator of the age of an YSO.  There is a positive 
trend observed between the polarization and H-K colour.  The sources 
that display the highest levels of polarization have H-K$>$2.5, the 
sources that have polarizations between 1.5\% and 5\% have the 
greatest range of H-K colours.  Of the two sources that show 
polarizations of less than 1.5\% one has H-K=2.1, and the other has 
the lowest H-K colour of any of the sample (H-K $\approx$ 0.3). 
 
In figure 14 the near-IR colour-colour diagrams for the $\rho$ Oph 
sample sources are presented with K band polarization measurements. 
The open triangles indicate sources with polarization less than 1.5\%, 
filled squares represent sources with polarization between 1.5\% and 
5\%, and open squares indicate sources with polarization greater than 
5\%.  The colour-colour plots indicate that the polarization 
correlates with near-IR excess emission, as measured by the H-K and 
K-L colour index. All sources with K-L $\geqslant$ 1.5 have 
$\it{P}$$_K$ $\geqslant$ 1.5\%.  Aside from DOAR25 all stars with H-K 
$\geqslant$ 1.2 also have $\it{P}$$_K$ $\geqslant$ 1.5\%. 
 
\section{Models} 
 
The shadow.f code (Lucas et al. 2004; Lucas 2003) can be used to 
represent a YSO with either 2-D, axisymmetric model or a 3-D, 
non-axisymmetric model.  The distribution of matter is based on a 
simple star, disc, envelope system.  The grains in the system are 
oblate spheroids; their appearance does not change with rotation about 
the short axis.  It is assumed that the short axis of the grains is 
perfectly aligned with the magnetic field and that the grains spin 
about this axis. Large numbers of photon packets are generated at the 
surface of the protostar and these then move through the system, 
suffering modifications to their Stokes vectors due to dichroic 
extinction and scattering events. 
 
The model results produced by the shadow.f code are output as Stokes 
I, Q and U images for comparison with the observational data.  The 
pixel scale applied to the Stokes parameters is the same as that for 
the UKIRT polarization data; 0.286 arcseconds pixel$^{-1}$ or 0.143 
arcseconds pixel$^{-1}$ with the magnifier in place (the distance to 
$\rho$ Oph is assumed to be 160 parsecs).  The Stokes I, Q, U are 
convolved using a Moffat profile with the form 
 
\begin{equation} 
\frac{1}{\left( 1+ \left( \frac{R}{\alpha} \right)^2 \right)^\beta} 
\end{equation} 
 
\noindent{where $\alpha$ and $\beta$ are two constants determined from 
the point-spread-function (PSF) of a standard star.  The POLPACK 
package is then used to produce the degree of polarization map and the 
vector catalogue.} 
 
\subsection{Input Parameters} 
 
The shadow.f code uses numerous input variables, which are listed in 
Table 4.  The recommended minimum number of input photons is 100,000. 
The radius of the system (i.e. the large scale envelope), the outer 
radius of the disc, the base radius and the opening angle of the 
cavity  manipulate the physical dimensions of the system. All four 
variables can be estimated from the observational results.  The polar 
and azimuthal viewing angles can also be estimated from the data.  A 
polar viewing angle of 0 or 180 degrees means that the observer is 
looking towards the pole of the object, an angle of 90 degrees 
provides an edge on view of the system. The density coefficient and 
the vertical density gradient are used to manipulate the structure of 
the envelope and its optical depth. 
 
\subsection{Density Distribution} 
 
The code has been adapted so that it can use two different envelope 
density equations.  The first is an empirical distribution that uses a 
power law index to fix the vertical density gradient. 
 
\begin{equation} 
\rho_{env}=CR^{-3/2}\left(\frac{1}{\mu^k+0.05}\right) 
\end{equation} 
 
\noindent{where $\it{C}$ is a free parameter whose value is is the 
density coefficient in units of kg m$^{-3/2}$). In most cases the line 
of sight to the protostar passes through the envelope but not the 
accretion disc, so $C$ is simply proportional to $\tau_K$, the optical 
depth at K band, which is one of the parameters listed in Table 4. 
$\it{R}$ is the 3-dimensional radius for the natural system of 
cylindrical polar coordinates defined by the disk rotation axis 
($\it{R}$ = $\sqrt{r^2+z^2}$); the azimuthal coordinate is $\Lambda$), 
$\mu$=z/r, and k is the power law index (increasing the value of 
$\it{k}$ increases the degree of flattening of the envelope).} 
 
The second is based on the density distribution of Terebey, Shu and 
Cassen (1984), 
 
\begin{equation} 
\rho_{env}=CR^{-3/2}\left(1+\frac{\mu}{\mu_0}\right)^{1/2}\left(\frac{\mu}{\mu_0}+\frac{2r_c\mu_0^2}{R}\right)^{-1} 
\end{equation}		 
 
\noindent{where $\it{r}$$_c$ and $\mu_0$ are as described in that 
paper. Models that employ Equation 5 do not use the power law index, 
$k$, that is listed in Table 4 but replace this parameter with the 
``centrifugal radius'' $r_c$.} 
 
\subsection{YLW16A} 

The bipolar nature of this sytem can be inferred from the polarization maps and
polarized flux maps in figure 4(b-c). However, the image in figure 4(a) does not resolve 
resolve the structure, so we illustrate it better with additional data shown in
figure 15(a-b). Figure 15(a) shows a NICMOS image at 0.15 arcsec resolution
from the HST archive. The central regions of the system are seen to be double peaked.
Figure 15(b) shows a deep ground based image of a wider area reproduced from Lucas \& 
Roche (1998), with slightly higher spatial resolution than the polarimetric
data in figure 4. This image shows the arrowhead structure mentioned in
$\S$4.4.

\subsubsection{Model Parameters} 
 
Models with an evacuated cavity along the system axis (the default 
assumption) failed to qualitatively reproduce either the arrowhead 
structure or the double-peaked nature of the core.  
We therefore attempted to reproduce these features by 
modelling YLW16A as a bipolar source with a dusty jet inside the 
bipolar cavity.  Dust was introduced into both lobes of the bipolar 
cavity.  Low density dust in the western lobe of the cavity (the part 
tilted away from the line of sight) reflects light from the protostar 
that would otherwise escape the system unseen, and serves to increase 
the prominence of the western peak. High density dust in the eastern 
lobe is used to increase the obscuration of the protostar and the 
eastern part of the reflection nebula so that the jet feature produced 
by dust in the western lobe appears relatively bright.  The eastern 
lobe is slightly curved in the parabolic sense with cavity walls at: 
 
\begin{equation} 
r = R_c + |z| tan(\theta_c) \left(\frac{R_c}{R} \right)^{cc} 
\end{equation} 
 
\noindent{where $\theta_c$ is a parameter equivalent to the opening 
angle close to the disc plane, $\it{R}$$_c$ is the radius of the 
cavity in the disc plane, and cc is the parabolic curvature parameter. 
The dust in the eastern lobe is located throughout the cavity from the 
stellar surface to a radius, $\it{R}$=1000 AU, but only for azimuthal 
angles $\left| \Lambda \right|$ $<$ 100$^{\circ}$ ($\Lambda$=0 is the 
azimuth of the line of sight).  This increases the extinction toward 
the inner regions of the eastern lobe, while allowing some light 
through to illuminate the large scale reflection nebulosity that is 
observed. In order to better reproduce the narrow jet-like feature 
seen in figure 15) in the western lobe, the cavity on that side  
has a narrow cylindrical structure described by $\it{r}$ $<$ $\it{R}$$_c$  
and dust fills the region at:} 
 
\begin{equation} 
R_{CD}<R<270 AU. 
\end{equation} 
 
\noindent{with density $\it{\rho_{cav}}$ = 
3$\times$10$^{-13}\it{G_{CD}}$. The density in the eastern lobe of the 
cavity is:} 
 
\begin{equation} 
\rho_{cav}=3\times 10^{-13}G_{CD2} \left(1+ \left(\frac{R}{R_c} 
\right)^2 \right)^{-1} kg m^{-3} 
\end{equation} 
 
\noindent{where G$_{CD}$ and G$_{CD2}$ are free parameters, with CD referring 
to cavity dust.} 
 
The Moffat profile used to convolve the Stokes elements was described 
by $\alpha$ = 6.49 and $\beta$ = 3.35.  These values were determined 
from the point spread function (PSF) for the point-like young star 
YLW16B imaged on the same night as YLW16A. 
 
\subsubsection{Model Fit}
 
The final model provides a quantitative fit to the position angle of 
the polarization vectors and degree of polarization over the core, 
which was assessed in a 2 arcsecond aperture and is centred at the 
point between the two peaks.  In addition it provides a quantitative 
fit to the maximum polarization assessed in a 0.5 arcsecond aperture 
centred approximately 3.5 arcseconds from the core. 

\begin{figure} 
 \vspace{-4.5cm}\hspace{-8cm}\includegraphics[scale=1.0,angle=0]{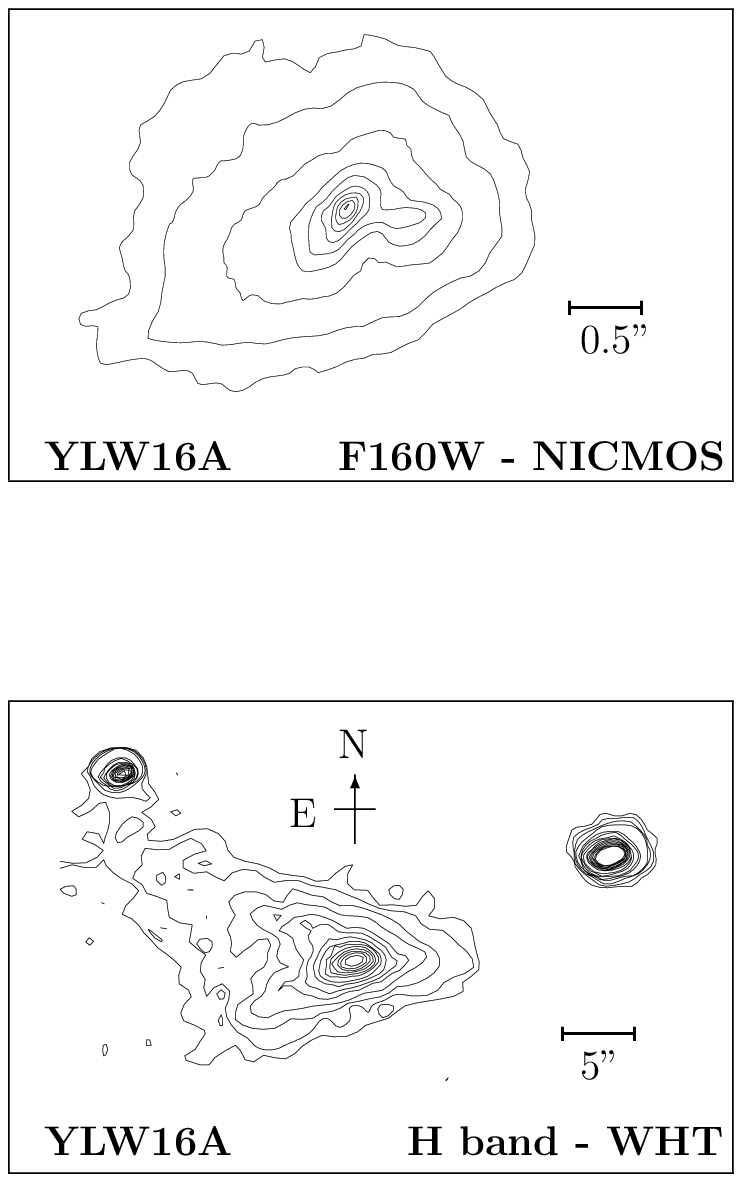}\vspace{-1cm} 
\vspace{-12cm}
 \caption{Images of YLW16A. (top) NICMOS image in the F160W filter, showing the bipolar nature 
of the system. We interpret the small scale narrow extension in the western lobe as 
reflection from a dusty jet. Data from the HST archive (programme xxxx, PI S.Terebey). 
(bottom) Wider field H band image reproduced from Lucas \& Roche 1998. This image shows the 
extended nebulosity and the arrowhead structure more clearly than the polarimetric data
in figure 4. The figures have the same orientation as the data in figures 1 and 4, i.e. north
is up and east is to the left.} 
\end{figure} 
 
\begin{figure} 
 \vspace{-5.1cm}\hspace{-7.3cm}\includegraphics[scale=1.0,angle=0]{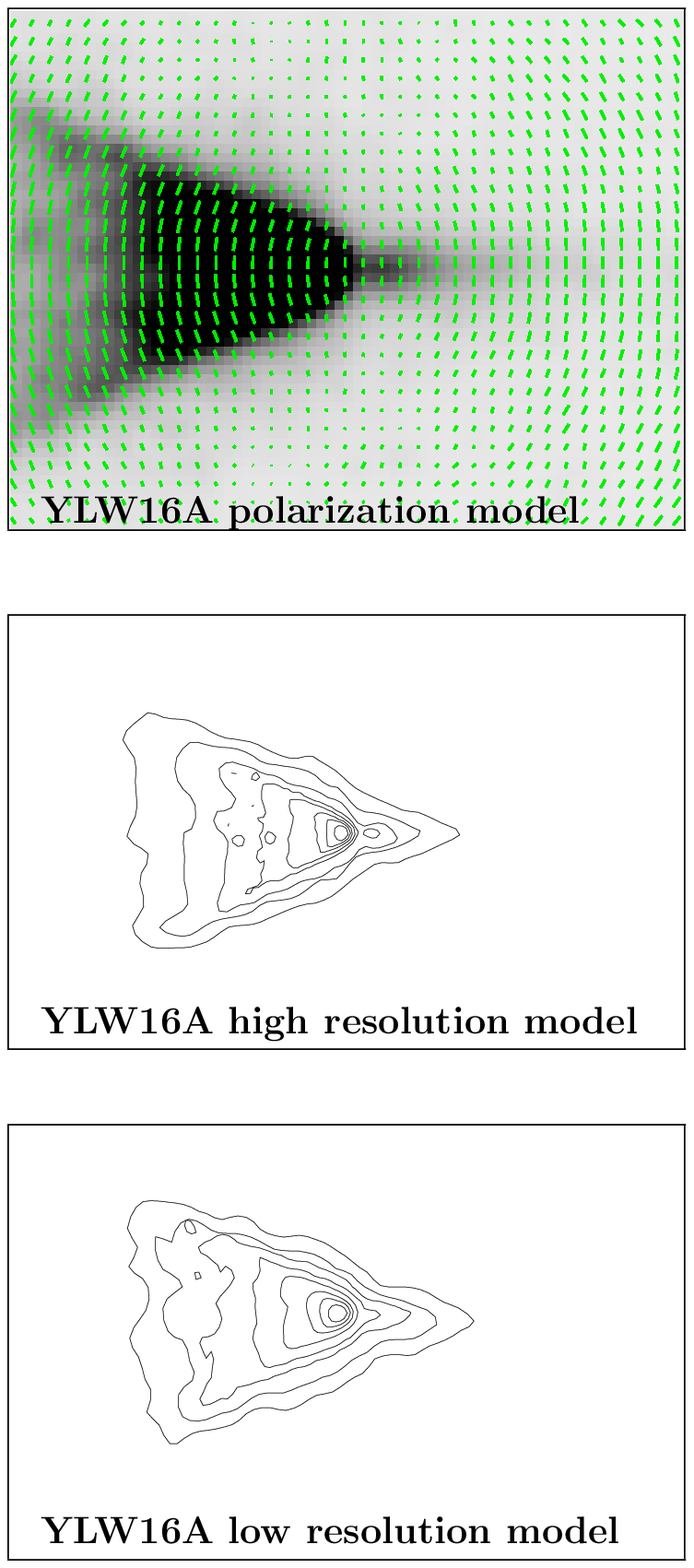}\vspace{-1cm} 
\vspace{-6cm}
 \caption{The model fit for YLW16A. (top) polarization vector map, smoothed to the approximate
resolution of the data in figure 4. The toroidal field structure reproduces the prominent 
polarization disc.(middle) flux distribution, shown at the approximate
resolution of the NICMOS image. Note that we do not attempt to reproduce the northwest-southeast
orientation of the inner contours seen in figure 15. (bottom) flux distribution shown at the 
approximate resolution of the ground based data. The figure has the same orientation as 
figure 15.}
\end{figure} 
 
The model fit to YLW16A is shown in figure 16, the fit parameters are 
summarised in table 4. 
 
The model provides a qualitative fit to the total flux distribution 
and the polarized intensity distribution.  By placing dust in both 
cavities it was possible to reproduce the second peak seen in the NICMOS 
data.  If dust was 
placed in only the western (right hand) lobe it was not possible to 
reproduce the second peak for any value of G$_{CD}$ investigated, 
hence the need for the dense dust in the eastern lobe.  The fit is 
achieved for values of the upper and lower density coefficients of 
G$_{CD2}$ = 6$\pm$1.5 and 0.01 $<$ G$_{CD}$ $<$ 0.1. We note that
the second peak is less prominent in the model than in the NICMOS
data, so we do not rule out the possibility of a second illuminating source
at that location.
 
\begin{table} 
\begin{minipage}{80mm} 
    \caption{Parameters for model fit to YLW16A at the K band.} 
    \begin{tabular}{@{}lc@{}} 
  \hline Parameter & Value\\ \hline System Radius & 3400 AU\\ Outer 
   Radius of Disc, $\it{r}_o$ & 100 AU\\ Envelope Optical Depth to protostar at 
   K, $\tau_K$ & 8.6\\ Power Law Index, $\it{k}$ & 0\\ Polar Viewing 
   Angle, $\theta$ & 66$^{\circ}$.4\\ Azimuthal Viewing Angle, 
   $\Lambda$ & 0$^{\circ}$\\ Base Radius of Cavity, R$_c$ & 25 AU\\ 
   Half-opening Angle of Cavity at base, $\theta_c$ & 25$^{\circ}$\\ 
   Grain Albedo & 0.4\\ Grain Axis Ratio & 1.015 $>$ gr $>$ 1.03\\ 
   Maximum Grain Size & 1.05 $\mu$m\\ Minimum Radius of Cavity Dust, 
   R$_{CD}$ & 25 AU\\ Lower cavity Density Coefficient, G$_{CD}$ & 
   0.1\\ Upper cavity Density Coefficient, G$_{CD2}$ & 6\\ Parabolic 
   Curvature Parameter, cc & 0.2\\ \hline 
\end{tabular} 
\end{minipage} 
\end{table} 
 
The best fit parameters are listed in Table 4. By comparing the 
prominence of the eastern and western lobes as seen in the ground based data
in figure 4 and figure 15, the polar viewing angle 
was constrained to 60$^{\circ}$ $\leqslant$ $\theta$ $\leqslant$ 
73$^{\circ}$.  In the final model a polar viewing angle of $\theta$ = 
66$^{\circ}$ (cos($\theta$)=0.4) was found to provide the best fit to 
the data.  The opening angle of the cavity for the final model was 
$\theta_c$ = 25$^{\circ}$ (from the disc axis to the cavity wall), and 
the value of $\it{R}$$_c$ was 25 AU. 
 
The Terebey, Shu \& Cassen envelope density equation (5) and the 
empirical envelope density equation (4) were both investigated.  Both 
equations were capable of producing similar results.  The final model 
used the empirical equation with a power law index of $\it{k}$ = 0. 
More positive values of the power law index failed to reproduce the 
double peaked structure.  The Terebey, Shu \& Cassen density equation 
was capable of providing a qualitative fit to the arrowhead structure 
for a centrifugal radius of $\it{R}$$_c$ $\leqslant$ 100 AU, which 
approximates to the simple $R^{-3/2}$ distribution as the final model 
as ${R}_c \rightarrow 0$. 

The total optical depth to the protostar, including the contributions 
of the cavity dust and the envelope, is $\tau_K=18$, so the protostar 
itself is entirely obscured from our line of sight. The optical 
depth must be lower along other lines of sight to permit the illumination 
of the large reflection nebula, which is why the cavity dust in the  
eastern lobe is restricted to a certain range of azimuthal angles, as noted 
earlier. 
 
The optical depth through the envelope to the protostar in the K band  
was investigated for values of the envelope density coefficient $C$ that  
correspond to optical depth 4.6 $\leqslant \tau_K \leqslant$ 10.3 in the models 
(neglecting the optical depth of the dust inside the cavity).  For the 
preferred range of polar viewing angles it is possible to reproduce the double 
peaked structure for an envelope optical depth of $\it{\tau_K}$ = 
8.5$\pm$0.4.  The reproduction of the arrowhead structure, without 
reproducing the double-peaked structure, increases the range of 
acceptable optical depths to 8.0 $\leqslant$ $\tau_k$ $\leqslant$ 9.5. 
Decreasing the optical depth of the envelope
below this range resulted in a total flux distribution
that was dominated by the flux from the western peak and a broadening 
of the lower lobe, thus the arrowhead shape disappeared.  Increasing 
$\tau_K$ above this range resulted in a less prominent western peak 
and no distinct arrowhead structure. 
 
In the final model the dust grains were aligned with a toroidal 
magnetic field.  An axial magnetic field was able to reproduce the 
polarization vector structure in the extended nebula, but failed to 
reproduce the core polarization structure. The outer and inner 
polarization vector structure is quite successfully reproduced by either a 
toroidal field or simply a uniform interstellar field oriented 
parallel to the disc plane. The actual field structure may well be more
complicated than this: as noted in $\S$4.4 the the inner contours of the
flux distribution have a northwest-southeast orientation (see figure 4a and 
figure 15, top panel) which is inclined with respect to the larger scale
east-west structure. This northwest-southeast orientation of the inner regions
is also in the polarization vectors. Our description of the field structure 
as toroidal is only approximate, given that the orientation of the disc
plane is clearly somewhat uncertain. We note that while field lines approximately
parallel to the disc plane appear to be required on scales of a few hundred AU, 
the field structure may of course become more axial on larger scales. 
The possible existence of a second protostar at the location
of the second peak in the NICMOS data is not expected to significantly
influence the polarization maps, since the 0.5 arcsec from the principal
flux peak is unresolved in the ground based data.
 
The observed polarization pattern and polarization levels associated 
with YLW 16A can be well-modelled using dust grains with a maximum 
size of 1.05 $\mu$m.  Smaller maximum grain sizes were considered, 
however they resulted in core polarizations that were higher than 
required. The grain axial ratio adopted was 
1.015$_{-0.005}^{+0.015}$. Grain axial ratios above or below this 
range led to core polarizations that were significantly higher or 
lower than required, respectively. Note that since these grains are 
almost spherical, the magnetic field direction direction only affects 
the core polarization and has negligible effect on the other model 
parameters. It is of course possible that the core polarization is due 
to grains which are highly aspherical but only weakly aligned with the 
magnetic field. 
 
\subsection{WLY54} 
 
\subsubsection{Model Parameters} 
 
In Figure 5 there is an oblique angle between the normal to 
the polarization disc plane and the extended nebula. We suggest that this  
could be explained if WLY54 is a bipolar nebula with a foreground extinction  
cloud. The vectors reveal the possible existence of a small counterlobe to the 
southwest of the core. The position angle of the polarization disc 
relative to the direction of the extended nebula would be explained if there 
were a source of extinction to the north of WLY54 that obscures part 
of the nebula.  Further support for the bipolar nature of WLY54 comes  
from the east - west orientation of the degree of polarization map, which  
shows a low polarization region running north - south, with two higher  
polarization regions to either side.  The bipolar nature of WLY54 is further 
supported by the appearance of the less prominent (western) lobe in an 
H-K colour map  (not shown) which appears to be redder than the more 
prominent (eastern) lobe. However, there is no direct evidence for the 
foreground extinction cloud that is used in the model to reproduce the 
unusual total flux and polarization structures.  If there is no 
foreground extinction cloud then something else must be responsible 
for the misalignment between the disc axis and the extended 
nebulosity. Two possible explanations are that the orientation of the extended 
nebulosity is influenced by source movement through a dense medium, or  
extinction within the cavity. Alternatively, the polarization disc 
may not be a feature of the system at all, but might simply by due to 
foreground dichroic extinction if the polarization intinsic to the WLY54  
system is very low in that region. We note that most of the sources in this  
$\rho$ Ophiuchi sample have redder colours than the sources in the Taurus 
sample of Whitney et al.(1997), which implies higher foreground extinction.  
This is particularly obvious for the Class II systems, which have less  
intrinsic extinction. 
 
The location of the foreground extinction cloud along the line of 
sight is defined by somewhat arbitrary limits in the model (see Table 5).  
The structure of the extinction cloud in the plane of the sky is described 
by a gradient that begins 200AU to the north of the core.  Two 
extinction gradients were investigated.  The first is an exponential 
profile described by 
 
\begin{equation} 
\rho_{screen}=C\tau_s e^{\Lambda} 
\end{equation}									 
 
\noindent{where $C$ is a constant (here given the value 4.55 
$\times$ 10$^{-15}$  kg m$^{-3}$), $\tau_s$ is a variable that can be 
used to manipulate the optical depth of the extinction cloud, and 
$\Lambda$ describes the azimuthal angle.  The second is a linear 
profile} 
 
\begin{equation} 
\rho_{screen}=C\tau_s rsin\Lambda 
\end{equation}									 
 
\noindent{where the constant $C$ is here given the value 3 $\times$ 
10$^{-29}$ kg m$^{-4}$, and $\it{r}$sin($\Lambda$) is the distance 
north of the protostar.} 
 
It was assumed that the density distribution of the envelope is 
adequately described by the empirical density equation (10). 
 
The Moffat profile values are determined from the PSF for the 
point-like YSO YLW13A imaged during the same observing run as WLY54 
($\alpha$ = 3.71 and $\beta$ = 3.49).

\subsubsection{Final Model} 
 
The parameters that provide the fit to the H band data are summarized 
in Table 5 and the fit is illustrated in figure 17. 
 
The model provides a quantitative fit to the maximum degree of 
polarization in WLY54 (evaluated in a 0.5 arcsecond aperture), and the 
position angle and degree of polarization of the polarization disc 
(evaluated in a 2 arcsecond aperture). 
 
\begin{figure} 
 \vspace{-1.9cm}\hspace{-1.3cm}\includegraphics[scale=0.6,angle=0]{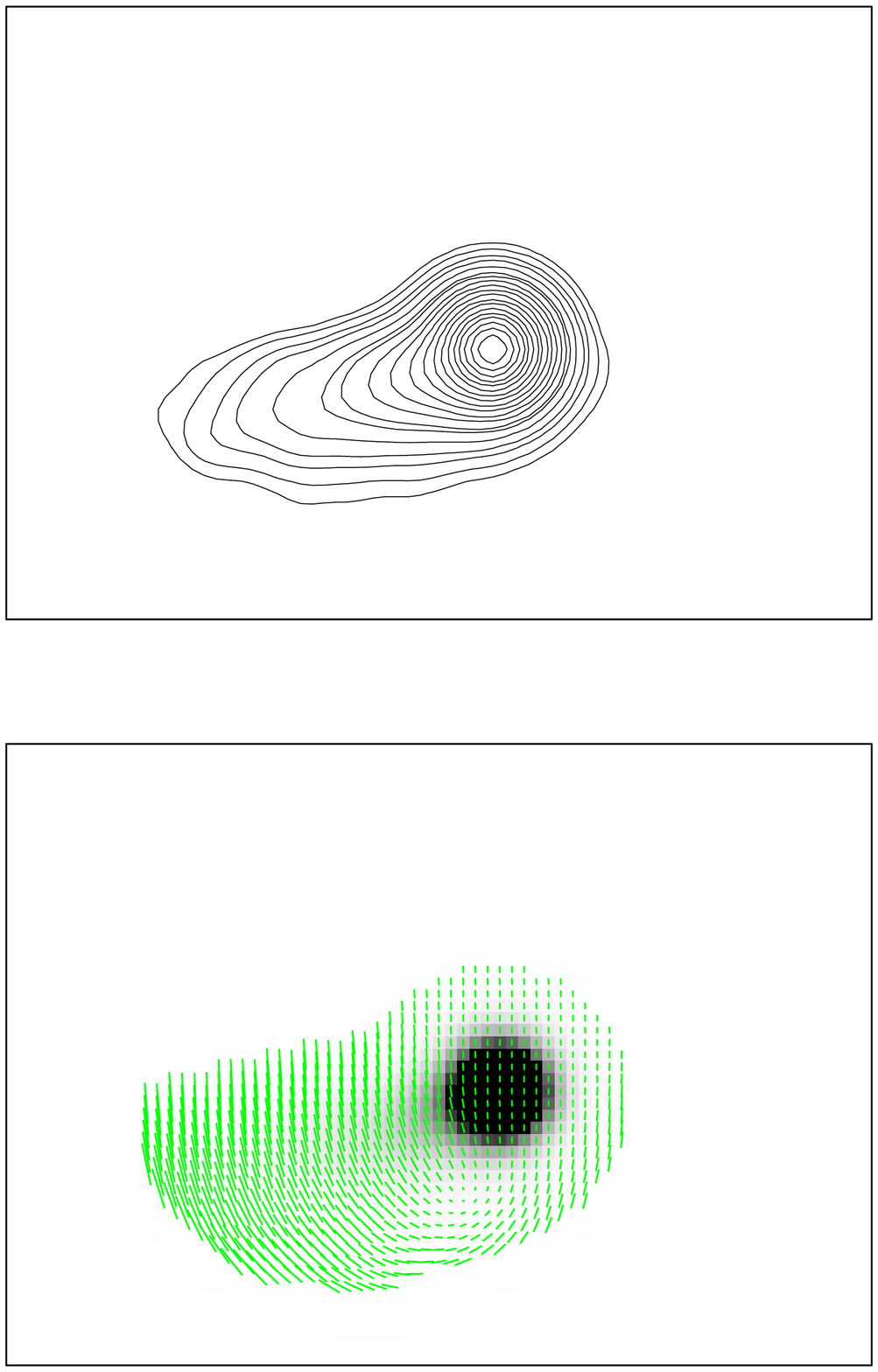}\vspace{-1cm} 
 \caption{ The model fit for WLY54 (top) flux distribution, and (bottom) 
 polarization vector map.} 
\end{figure} 
 
The model provides a qualitative fit to the shape of the cometary 
nebula as revealed by the total flux distribution.  In addition the 
model successfully reproduced the angle between the orientation of the 
polarization disc and the direction of extension.  These features are 
reproduced by introducing a foreground extinction cloud to the north 
of the system.  The structure of the total flux distribution was best 
reproduced when the influence of the foreground extinction cloud was 
described by the linear profile defined in equ.10.  The exponential 
profile did not have a significant influence on the structure of the 
total flux of WLY54.  It was also found that parabolic cavity walls 
produced a better fit than conical cavity walls.  The parabolic cavity 
walls are defined by equ.6, which introduces a curvature parameter cc 
to the conical profile. 
 
The cometary appearance of WLY54 suggested a polar viewing angle that 
minimises the amount of counterlobe that was visible.  The possible 
presence of the foreground extinction cloud in the system means that 
the polar viewing angle is not tightly constrained since acceptable 
results were found for $\it{\theta}$ = 45$^{\circ}$.6$_{-9}^{+14}$. 
The adopted polar viewing angle for the final model provided the best 
fit to both the total intensity distribution and the polarization data.

A suitable qualitative fit to the total flux distribution was found 
for a power law index of $\it{k}$ = 1.0$\pm$0.4 (see 
Equ.4). Increasing k reduces the prominence of the western lobe for 
the chosen optical depth. 
 
To produce a reasonable fit to the radial intensity profile of the H 
band data requires an optical depth in the range 5.4 $<$ $\tau_H$ $<$ 
6.5.  Optical depths outside this range either resulted in a total 
flux distribution that had too steep an intensity profile, for 
$\tau_H$ $<$ 5.4, or too shallow an intensity profile, for $\tau_H$ 
$>$ 6.5. The main 
problem encountered when trying to further constrain the optical depth 
was the extremely low levels of core polarization shown by WLY54 at 
all wavelengths.  The best fit to the radial intensity profile is 
achieved for an optical depth closer to $\tau_H$ = 6.5, whereas the 
best fit to the polarization data is achieved for an optical depth of 
$\tau_H$ = 5.5. 
 
The dust grain albedo was treated as a free parameter.  The grains 
were a mixture of silicates and amorphous carbons.  For the polar 
viewing angle adopted, $\it{\theta}$ = 45$^{\circ}$.6$_{-9}^{+14}$, 
the best fit to the data was found for a grain albedo of approximately 
0.45 at H. 
 
The position angle of the polarization disc and the outer polarization 
vector structure were reproduced by dust grains aligned with a 
toroidal magnetic field.  An axial magnetic field led to the position 
angle of the polarization vectors over the core being at 90$^{\circ}$ 
to the requirement due to dichroic extinction. We note that a return 
to an axial field structure is of course possible on larger spatial  
scales, and is expected from basic star formation theory. To fit the core 5.3\% 
polarization observed at UKIRT in the H band with the adopted magnetic 
field structure and maximum grain size of 0.35 $\mu$m required that 
the grain axial ratio was not larger than 1.015.  Increasing the grain 
axial ratio resulted in core polarization levels that were 
significantly larger than required.  A fit to the degree of core 
polarization and position angle of the polarization disc was also 
found using larger grains. However, the models with larger grains were 
not able to reproduce the high polarization observed over the more 
extended envelope. 
 
\begin{table} 
\begin{minipage}{80mm} 
    \caption{Parameters for model fit to WLY54 at the H band.} 
    \begin{tabular}{@{}lc@{}} 
  \hline Parameter & Value\\ \hline System Radius & 1560 AU\\ Outer 
   Radius of Disc, r$_o$ & 100 AU\\ Optical Depth to protostar at H, 
   $\tau_H$ & 5.95$\pm$0.55\\ Power Law Index & 1.0$\pm$0.4\\ Polar 
   Viewing Angle, $\theta$ & 45$^{\circ}$.6\\ Azimuthal Viewing Angle, 
   $\Lambda$ & 0$^{\circ}$\\ Base Radius of Cavity, R$_c$ & 30 AU\\ 
   Half-opening Angle of Cavity at base, $\theta_c$ & 23$^{\circ}$\\ 
   Parabolic Curvature Parameter, cc & 0.2\\ Grain Albedo & 0.45\\ 
   Grain Axis Ratio & 1.015\\ Maximum Grain Size & 0.35 $\mu$m\\ Inner 
   Radius of Extinction Cloud & 2180 AU\\ Outer Radius of Extinction 
   Cloud & 4680 AU\\ Density parameter of Extinction Cloud, $\tau_s$ & 
   30\\ \hline 
\end{tabular} 
\end{minipage} 
\end{table} 
 
Spherical grain models were capable of producing a reasonable 
approximation to the WLY54 data.  However, there are some key 
problems. The model fails to reproduce the polarization vector 
structure over the core region, and the degree of core polarization is 
approximately one tenth of the requirement.  The model does manage to 
reproduce the maximum polarization and provides a qualitative fit to 
the total flux distribution.  To fit the 5.3\% polarization requires 
an optical depth of $\tau$ = 8.7$\pm$0.6.  The polarization vector 
structure for an optical depth in this range is comparable to the 
data. However the model does not provide a good fit to the observed 
flux distribution. 
 
The density parameter of the foreground extinction cloud was 
investigated across a broad range of values, 5 $\leqslant$ $\tau_s$ 
$\leqslant$ 50, as it was not known how its presence would affect the 
system.  Increasing the value of $\tau_s$ causes a reduction in the 
flux from both lobes to the north of the core.  The 
final model required that the density parameter was $\tau_s$ = 
30$\pm$3 to reproduce the structure of the total flux distribution in 
the eastern lobe.  For the range of optical depths that were capable 
of reproducing the total flux distribution in the eastern lobe the 
foreground extinction cloud completely suppressed the appearance of 
the western lobe at all polar viewing angles for k $\geqslant$ 1.0. 
For $\it{k}$ = 0.6 and $\theta$ = 53$^{\circ}$.1 a small western lobe 
is visible. 
 
Increasing the cavity curvature parameter results in a narrowing of 
the extended nebulosity, a qualitative fit to the total flux 
distribution is found for cc = 0.2 when combined with the influence of 
the foreground extinction cloud. 
  
\subsection{EL29} 
 
\begin{figure} 
 \vspace{-1cm}\hspace{-1.3cm}\includegraphics[scale=0.6,angle=0]{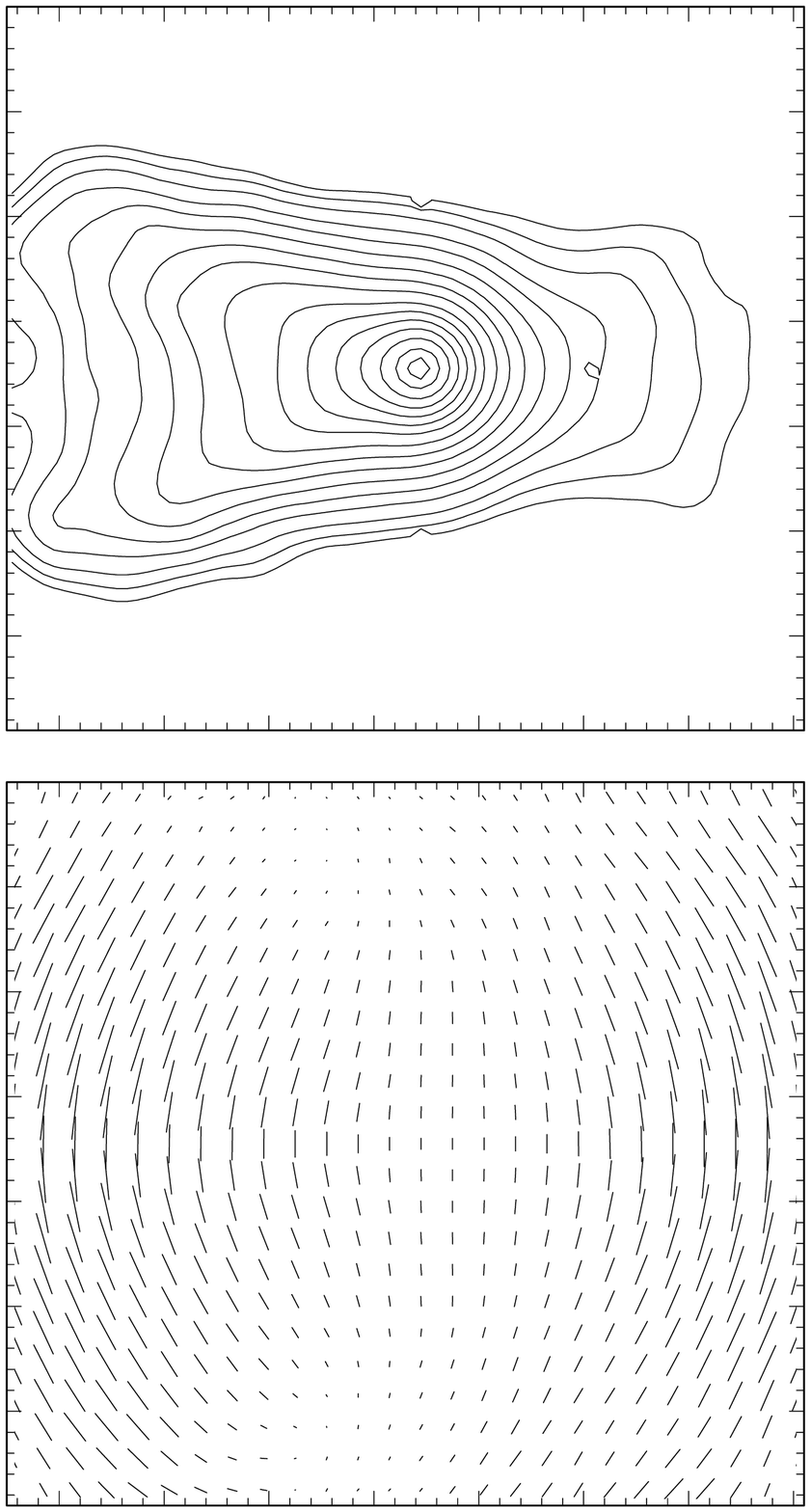}\vspace{-1cm} 
 \caption{The model fit for EL29 (a) polarized flux distribution, and 
 (b) polarization vector map.} 
\end{figure}

\subsubsection{Model Parameters} 
 
EL29 was modelled as a bipolar source.  The density distribution of 
the envelope was assumed to be adequately described by the empirical 
density equation, equ. 4.  The structure of the cavity was assumed to 
be conical (tan($\theta_c$) = 
(($\it{R}$-$\it{R}$$_c$)/$\left|z\right|$).  The base radius and 
opening angle of the cavity were estimated from the polarized flux 
data for EL29. 
 
A system radius of 1035 AU was estimated from the extent of the degree 
of polarization map, due to the point-like appearance of the total 
flux distribution.  The point-like nature of the flux distribution 
makes it difficult to determine the inclination angle of the system. 
For this reason EL29 was investigated as both a pole-on and edge-on 
object.  An object that is near pole-on would have a more point-like 
structure similar to that seen in the data.  However, if EL29 is a 
source whose true nature is obscured either due to the faintness of 
the nebulosity or the limited sensitivity of the observing technique then it 
is also possible that EL29 is closer to edge-on. 
 
The radius of the inner accretion disc, $\it{R}$$_a$, was investigated 
for values between 50 and 600 AU, which is the range of possible 
values suggested by Ceccarelli et al. (2002). 
 
The dust grains are oblate spheroids that are perfectly aligned by a 
magnetic field.  The magnetic field was investigated with both an 
axial and toroidal structure.  The albedo of the grains was treated as 
a free parameter. 
 
The final Stokes images (I, Q and U) were convolved using a Moffat 
profile described by $\alpha$ = 4.08 and $\beta$ = 1.86, for the 
pole-on models.  The values of $\alpha$ and $\beta$ are determined 
from the PSF for the point-like UKIRT photometric standard star FS140 
(s587-t) imaged during the same observing run as EL29. 
 
\subsubsection{Final Model} 
 
The parameters for the final model are given in Table 6.  The 
polarized flux distribution and polarization vector maps for EL29 and 
the final model are shown in figure 18. The final model provides a 
quantitative fit to the degree of core polarization, the structure of 
the polarization vectors and the maximum polarization measured in each 
lobe.  A qualitative fit is provided to the polarized flux 
distribution.  The fit is achieved by modelling EL29 as a system that 
is neither pole-on nor edge-on and has an outer disc radius of 50 AU. 
 
The inclination of the system was investigated for polar viewing 
angles in the ranges 0$^{\circ}$ $<$ $\theta$ $<$ 60$^{\circ}$  
(near pole-on) and from 80$^{\circ}$ $<$ $\theta$ $<$ 90$^{\circ}$ (edge-on). 
The point-like nature of EL29 in the Stokes I image was not 
successfully modelled; instead the focus was turned to modelling the 
polarized flux distribution, the polarization vector structure and the 
fractional polarization.  The inclination angle of the system was determined from 
previous observations of EL29 that have indicated the presence of a 
flared disc with $i\leqslant$ 60$^{\circ}$ (Boogert et al. 2002).  Acceptable results 
for the polarized flux distribution and to the maximum polarizations 
in each lobe were found for an inclination angle of $\it{i}$ = 
53$^{\circ}$.1$_{-8}^{+7}$. 
 
The values of the cavity base radius and half-opening angle 
investigated were based on the structure of the polarized flux 
distribution.  A reasonable fit to the polarized flux distribution can 
be found for cavity base radii of and half-opening angles in the 
ranges 20 AU $\leqslant$ $\it{R}$$_c$ $\leqslant$ 30 AU and 
20$^{\circ}$ $\leqslant$ $\theta_c$ $\leqslant$ 25$^{\circ}$. 
 
The optical depth to the protostar was investigated between 4 
$\leqslant$ $\tau_K$ $\leqslant$ 11.  The final model requires that 
the optical depth is $\tau_K$ = 7.75$_{-1}^{+0.25}$. Optical depths 
outside this range were not able to reproduce a reasonable fit to the 
polarized flux distribution. 
 
\begin{table} 
\begin{minipage}{80mm} 
    \caption{Parameters of the model fit to EL29 at the K band.} 
     \begin{tabular}{@{}lc@{}} 
  \hline Parameter & Value\\ \hline System Radius & 1035 AU\\ Outer 
   Radius of Disc, r$_o$ & 50 AU\\ Optical Depth to protostar at H, 
   $\tau_K$ & 6.75\\ Power Law Index, k & 0.0\\ Polar Viewing Angle, 
   $\theta$ & 53$^{\circ}$.1\\ Azimuthal Viewing Angle, $\Lambda$ & 
   0$^{\circ}$\\ Base Radius of Cavity, R$_c$ & 20 AU\\ Half-opening 
   Angle of Cavity at base, $\theta_c$ & 20$^{\circ}$\\ Grain Albedo & 
   0.2\\ Grain Axis Ratio & 1.015\\ Maximum Grain Size & 0.70 $\mu$m\\ 
   \hline 
\end{tabular} 
\end{minipage} 
\end{table} 
 
The empirical density equation was used to determine the density 
distribution.  The power law index was capable of producing acceptable 
results for k = 0.0$^{+0.4}$.  Increasing the power law index 
(i.e. the vertical density gradient) resulted in a decrease in the 
prominence of the receding lobe.  A more severe flattening of the 
envelope was not able to reproduce the polarization structure of EL29. 
 
The position angle of the polarization disc was reproduced using dust 
grains that were aligned with a toroidal magnetic field. Grains 
aligned with an axial magnetic field were capable of reproducing the 
vector structure in the outer parts of the polarised flux image but 
they were not able to reproduce the position angle of the polarization 
disc. To fit the degree of core polarization, P$_K$ = 8.3\%, required 
that for a maximum grain size of 0.70 $\mu$m the grain axial ratio be 
not more than 1.03 for the adopted magnetic field structure.  The 
larger grain axial ratio models and the spherical grain models 
resulted in core polarizations that were too high or too low, 
respectively. 
 
The dust grains were a mixture of silicates and amorphous carbons. 
For the adopted system inclination the dust grain albedo needed to be 
$\omega_K$ $\leqslant$ 0.4.  The final model has an albedo of 
$\omega_K$ = 0.2, which provided the best fit to the polarization 
data.  Increasing the albedo results in an increase in the prominence 
of the western (receding) lobe. 
 
This model fails to reproduce the point-like flux distribution that we 
observed for EL29,  though we noted in $\S$4.8 that more recent 
observations have detected the faint extended  structure that our 
model predicts. One possible reason for the faintness of the extended 
nebulosity is that the adopted envelope structure assumes that there 
is an axisymmetric distribution of material throughout. However, the 
strongly peaked structure of EL29 suggests that there may be a 
``hole'' in the material along the line of sight to the protostar.

\section{Conclusions} 
 
The near-infrared linear polarization data for a sample of young 
stellar objects in the $\rho$ Ophiuchi star-forming region have been 
analysed.  The majority of the objects were spatially unresolved. 
Five of the objects were clearly associated with extended nebulosity: 
two of these are bipolar nebulae and three have cometary morphologies. 
 
The extended objects have centrosymmetric vector patterns with a 
polarization disc over the core. A few of the unresolved objects have 
polarizations that were too small to detect, with upper limits of 
1-2\%. 
 
The wavelength dependence of the degree of polarization suggests that 
dichroism is the dominant mechanism responsible for the polarization 
of light in the unresolved sources, which constitute the majority of the sample.  
(The alternative is that these sources have complicated but unresolved
polarization structures). In the 
envelopes of the extended sources the wavelength dependence indicates 
that scattering dominates the generation of polarized light. 
 
The results of the linear polarimetry survey of $\rho$ Oph were 
compared to the infrared evolutionary status of the objects.  It was 
found that: 
 
\begin{itemize} 
 
\item{The distribution of the material around the sample sources does 
not seem to be strongly correlated with the evolutionary scheme 
determined by the infrared spectral energy distributions.  Point-like 
structures were seen for objects with Class I, Class II and Class III 
designations.  Five objects were identified as being associated with 
extended nebulosity. The majority of the extended objects were Class I 
but one had a Class II designation.} 
 
\item{The near-infrared colour-colour diagrams show that the degree of 
core polarization is correlated with the H - K and K - L colours.} 
 
\item{A weak correlation is observed between the size of the extended 
nebulae and the core polarizations.} 
 
\item{There is a positive correlation between the degree of core 
polarization and the evolutionary stage indicated by the infrared 
spectral index.  Typically, Class I objects show core polarizations in 
the range 2\% $<$ P$_K$ $<$ 16\%, whereas core polarizations in the 
range 1\% $<$ P$_K$ $<$ 6\% are seen for the Class II objects, and the 
only Class III object in the sample shows PK $<$ 1\%. It should be noted 
that the SED-based assignation of classes is influenced by system  
inclination as well as evolutionary status.} 
 
\item{A weaker positive correlation between the maximum polarization 
assessed over the envelopes and the infrared spectral index was also 
observed.} 
 
\item{Generally, redder sources are associated with higher degrees of 
core polarization.  Those objects that have P$_K$ $\geqslant$ 5\% have 
(H - K) colours greater than 2.5. Objects with polarizations in the 
range 1.5\% $>$ P$_K$ $>$ 5\% have the greatest range of (H - K) 
colours (1 $<$ (H - K) $<$ 4.4), and the two objects with P$_K$ 
$\leqslant$ 1.5\% have (H - K) $<$ 2.2 (the object with the lowest H-K 
also has the lowest P$_K$).} 
 
\end {itemize} 
 
Similar correlations have been previously identified for young stellar 
objects in the Taurus region (Whitney, Kenyon \& Gomez 1997).  The 
results presented in this work for $\rho$ Oph and by previous authors 
for Taurus reveal that the polarization data appears to support the 
evolution of circumstelar matter predicted in models of low-mass star formation. 
 
The shadow.f code was used to model three Class I objects in $\rho$ 
Oph that represented a cross-section of the morphologies observed. 
The first was the bipolar object YLW16A, the second was the cometary 
object WLY54, and the third was the point-like object EL29.  In each 
case a fit to the polarization data was found by assuming that the 
objects could all be modelled as a bipolar system. 
 
The point-like source EL29 displays a bipolar polarization pattern 
across the image profile. This was assumed to be the result of diffuse 
bipolar nebulosity that was only detectable in polarized light. The presence of 
diffuse nebulosity has been confirmed by more recent observations. Our 
model fails to reproduce the point-like total flux distribution but 
successfully fits the polarized flux distribution and the polarization 
vector structure. 
 
The results of the modelling revealed that to reproduce the observed 
total flux distributions and polarization data there were several 
properties common to the three systems investigated.  These properties 
are: 
 
\begin{itemize} 
 
\item{The dust grains that are responsible for the polarization of the 
light are required to be smaller than the wavelength of observation. 
The maximum grain size was found to be 1.05 $\mu$m.} 
 
\item{The dust grains are either near-spherical or very weakly 
aligned. More strongly dichroic grains can only be present if the field 
structure varies along the line of sight to the protostar in a manner 
that weakens the measured dichroic extinction effects. 
Our models assume perfect grain alignment with the magnetic 
field. We find that grains with an axial ratio of more than 1.03 led 
to significantly higher polarizations than the data.  In addition, 
grains with axial ratios of 1.1 were found to result in little spatial 
variation in the polarization across the source for parameters that 
were capable of producing acceptable fits to the flux distributions 
investigated.} 
 
\item{Spherical grain models were successful at reproducing the flux 
distributions and polarization vector patterns, but they could not 
successfully reproduce the fractional polarizations measured in the 
cores of the YSOs.  Typically, the core polarization provided by a 
spherical grain model was a tenth of the observed value.} 
 
\item{Given that some dichroic extinction is required to reproduce the 
core polarizations, a toroidal magnetic field structure is necessary 
in the central regions of the nebula in  order to reproduce the 
orientation of the polarization disc parallel to the accretion  disc.} 
 
\end{itemize} 
 
\section*{Acknowledgments} 
 
        We wish to thank the staff of UKIRT, which is operated by the 
Joint Astronomy Centre on behalf of the UK Science and Technology 
Facilities Council (STFC). STFC has taken over the former astronomy 
functions of PPARC, the Particle Physics  and Astronomy Research 
Council. This research was supported by a PPARC PhD studentship 
awarded to AFB. PWL is supported by the STFC via an Advanced 
Fellowship at the University  of Hertfordshire.

\bsp 
 
\label{lastpage} 
 
\end{document}